 \pdfoutput=1
   \documentclass[fleqn,usenatbib,usedcolumn]{mnras}
   \usepackage[british]{babel}             
   \usepackage{txfonts}                    
   \let\la=\lesssim 
   \usepackage[T1]{fontenc}                
   \usepackage{graphicx}                   

\title[Environment of compact galactic nuclei]{Multi-phase environment of compact galactic nuclei:
the role of the Nuclear Star Cluster} 

\author[R\'o\.za\'nska et al.]
  { A.~R\'o\.za\'nska,$\!^1$  D. Kunneriath,$\!^{2,3}$
  B.~Czerny,$\!^{4,1}$  T.$\;$P. Adhikari,$\!^1$  and V. Karas$^3$\\ 
  $^1$Nicolaus Copernicus Astronomical Center, Polish Academy of Sciences, Bartycka 18,  00-716 Warsaw, Poland \\
  $^2$National Radio Astronomy Observatory, 520 Edgemont Road, Charlottesville 22903, VA, USA \\
  $^3$Astronomical Institute, Academy of Sciences, Bo\v{c}n\'{\i} II 1401, CZ-141\,00 Prague,
   Czech Republic\\
  $^4$Center for Theoretical Physics, Polish Academy of Sciences, Al. Lotnikow 32/46, 02-668 Warsaw, Poland}

\date{Accepted for publication in MNRAS}

\pubyear{2016}

\begin{document}
\label{firstpage}
\pagerange{\pageref{firstpage}--\pageref{lastpage}}
\maketitle

\begin{abstract}
We study the conditions for the onset of Thermal Instability in the innermost
regions of compact galactic nuclei, where the properties of the
interstellar environment are governed by the interplay of quasi-spherical
accretion onto a supermassive black hole (SMBH) and the heating/cooling
processes of gas in a dense nuclear star cluster. Stellar winds are the
source of material for radiatively inefficient (quasi-spherical,
non-magnetised) inflow/outflow onto the central SMBH, where a stagnation
point develops within the Bondi type accretion.  We study the local thermal
equilibrium to determine the parameter space which allows cold and hot 
phases in mutual contact to co-exist. We include the
effects of mechanical heating by stellar winds and radiative
cooling/heating by the ambient field of the dense star cluster. We consider 
two examples: the Nuclear Star Cluster (NSC) in the Milky
Way central region (including the gaseous Mini-spiral of Sgr~A*), and the
Ultra-Compact Dwarf galaxy M60-UCD1. We find that the two systems behave
in different ways because they are placed in different areas of
parameter space in the instability diagram: gas temperature vs.
dynamical ionization parameter. In the case of Sgr~A*, stellar heating prevents
the spontaneous formation of cold clouds. The plasma from stellar winds
joins the hot X-ray emitting phase and forms an outflow. In M60-UCD1 our
model predicts spontaneous formation of cold clouds in the inner part of
the galaxy. These cold clouds may survive since the cooling timescale is
shorter than the inflow/outflow timescale.
\end{abstract}


\section{Introduction}
The composition and state of interstellar gas and dust in galaxies vary across different morphological 
types and evolutionary stages  \citep{binney1998,swamy2005}. New stars form if the building
material is available but the process of  star-formation 
can be quenched by shock-heating and lack of supplies as galaxies evolve in the course of cosmological
 history \citep{kennicutt98,tacchella2015}. 
The environmental aspects are particularly complex in the central regions of galaxies, where 
strong tidal fields as well as effects of intense irradiation arise in
the sphere of influence of the supermassive black hole (SMBH). The latter  usually resides within a dense Nuclear Star Cluster (NSC) that may or may not be 
permeated by the dusty gaseous environment, depending on the specific conditions of a given source 
\citep{cole2015}. 

Our Milky Way's centre is a prominent example of a NSC that hosts both a SMBH and a large amount of material forming new stars even in its
recent past \citep{eckart2005,genzel2010}. 
The physics of NSCs in galaxies, namely, their origin, 
history and the composition of their 
interstellar environment  has a close resemblance on smaller scales to the properties of 
globular clusters \citep{degrijs2010}. Unlike nuclei of galaxies, typical globular clusters lack
SMBHs in their cores; also, they are generally unable to retain any significant amount of gas and 
dust \citep{frank76,moore11}. 
However, the category of Ultra-Compact Dwarf galaxies \citep{mieske2013,mieske2014} appears to 
be a suitable type of object that can 
help us to explore the physics and evolution of NSCs with a central SMBH, and to compare them 
with the properties of the NSC in our Galaxy.

The mechanism of cooling and heating of the interstellar medium (ISM) has been the object of detailed studies for some time
\citep{dyson1980,kennicutt1998}. 
The microscopic processes that govern the gas dynamics are well understood, but nevertheless 
we still lack a detailed understanding of the chemical composition of the dusty plasma and of the energy input from stars.  Such processes in galactic cores  are complicated further
  by the unusually high density of the NSC \citep{genzel2010,schodel2014,fritz2016}, the effect of accretion onto SMBH 
and possible feedback in the form of a jet \citep{moscibrodzka13,yuan2014}, vigorous starburst 
 activities, as well as  the role of interactions and mergers \citep{kennicutt2012}.  
 In this paper, we consider length-scales
 deep inside the NSC, where radiation and outflows  from stars as well as 
energetic emission  from the accretion flow, govern the physical properties of the system \citep{dopita2003}. 

The complexity of the gas dynamics in such a situation was described by \cite{silich2008} in their
 spherically symmetric picture. Very close to the  center the material flows in, and accretes onto the
 supermassive black hole. Far out the material has enough energy to escape in a transonic 
outflow, leading to a continuous loss of gas. If the radiative cooling is inefficient, the 
inflow/outflow regions are separated by a stagnation radius.  
The relevant solutions have been found by 
 \citet{quataert2004}. However, if plasma cooling is efficient,  instead of a simple stagnation 
 radius a whole region forms with no stationary solution \citep{silich2008}. A multi-phase medium 
 develops where cold (relatively dense) clouds can coexist together with the hot, diluted gas
  \citep{rozanska14}. Pressure is the same in both phases. Therefore, density and  temperature
can span a broad range of values. The fate of the cold phase is determined by the characteristic length 
\citep{field1965,defouw1970,begelman1990},
\begin{equation}
\lambda_{\rm F}=\left(\frac{\kappa T}{\Lambda n^2}\right)^{1/2},
\end{equation}
where $\kappa$ is the thermal conductivity of the medium, 
$T$ -- temperature, $n$ -- density number, and $\Lambda$ is 
the total cooling rate \citep{draine1984}.  Clouds of size less than $\lambda_{\rm F}$ tend to evaporate 
(heat conduction dominates), whereas larger clouds can be stabilized by radiative losses in the hot 
ambient medium (external heating vs.\ radiative cooling define the equilibrium of large clouds). 
Additional processes, such as turbulence and  magnetic fields can further modify 
the basic scenario (small-scale magnetic fields induce anisotropy by  inhibiting heat transport 
across the tangled field lines), but for the purpose of this paper we do not take them into account.
Here we consider a local development of Thermal Instability (TI), although larger-scale global 
instabilities may also be important for the evolution of the system as a whole \citep{ciotti2001}.

We study the above-mentioned type of TI, paying
attention to the description of the heating/cooling processes of matter with the 
use of the photoionization code {\sc cloudy} \citep{ferland1996,ferland2013}. 
A simple parametrization of the gas cooling rate from \cite{plewa1995} was employed 
by \citet{silich2008}, while we add a more 
detailed description of the system. In particular, we include the spectral shape of the incident
 radiation from NSC. Here, we consider two special cases. 
The first example concerns the dense NSC near the Galactic Center (GC) SMBH. 
In addition to our previous work \citep{czerny2013b,czerny2013c,rozanska14} where we have taken 
into account the role of radiation heating by accretion flow, now we also include the role 
of energy input by 
 radiation and wind outflows generated by hot stars. The second example falls within the category of 
 Ultra-Compact Galaxies (UCDs) as an extreme case of NSCs where the outer 
part of the galaxy is missing, possibly due to interaction. In particular, the proto-typical 
M60-UCD1 \citep{mieske2013,mieske2014} serves as an exploratory case. 

For their extreme properties -- compact dimensions and enhanced mass-to-light ratio -- 
UCDs fit somewhere in between the NSC of the Galactic centre on one side and globular clusters on the other. 
With respect to conditions for the TI, we expect different results for UCDs than for the GC
of the Milky Way because of their old stellar populations.
By examining a sample of UCDs in the Virgo Cluster, \citet{zhang2015} 
concluded that these are a qualitatively different type of object from luminous (otherwise ``normal'') globular clusters. \citet{seth2014} has shown that the dynamical 
mass of M60-UCD1's black hole reaches $2.1\times10^7M_\odot$ while its nuclear star-cluster half-light radius is only $\sim 24$~pc. 
We illustrate the main differences of the plasma distribution in these type 
of objects from the case of our GC. 

The structure of this paper is as follows. In Sec.~\ref{sec:mo} we introduce  the basic equations
and present  the full set-up of our model. 
All results are shown in Sec.~\ref{sec:res},  where we discuss two examples:
Sgr~A* in the centre of the Milky Way, and UCD1 associated with M60 galaxy.
Various aspects of the TI mechanism and potential ways to generalize our scenario are 
discussed in Sec.~\ref{sec:dis}. Finally, Sec.~\ref{sec:con} contains the main conclusions.

\section{Set-up of the Model}
\label{sec:mo}

\subsection{Conditions for Thermal Instability near SMBH}

Multi-phase ISM can form spontaneously by the irradiation-induced effects of TI. The radiation 
field may originate from NSC stars or from the central accreting SMBH. Only a
certain range of parameters allows TI to operate 
\citep[e.g.,][and further references therein]{cox2005}. The essential parameters of our 
system are the distance from the centre, the accretion rate onto the black hole, the corresponding
 radiative efficiency of accretion, and the energy input and  spectral distribution of stars in the NSC
\citep{field1965,krolik1981,rozanska96}. These complexities can be understood qualitatively: a 
clump of gas  cools on a time-scale $t_{\rm cooling}$,
 the exact value of which depends on the interplay of cooling mechanisms and their corresponding physical 
parameters such as: temperature, density, chemical abundances, and the dust content in ISM. 
Within the NSC, contact between the two phases (cold clumps embedded within a hot surrounding 
gas) is additionally affected by the continuous injection of fresh material in the form of 
stellar winds. This phenomenon of  
mass and energy input can lead to a stationary inflow/outflow solution, or a non-stationary 
process with a possibility of mass accumulation
 \citep{quataert2004,silich2004,silich2008}. 

We simplify the description of the dynamics and analyze just the direction of the thermal evolution
of the gas. Finally, we distinguish between two cases: the state of stable plasma in thermal equilibrium 
versus the cooling timescales relevant for plasma far from thermal equilibrium.
The plasma cools via free-free emission, and by atomic bound-free processes and  bound-bound transitions
 between different states. All these radiative cooling processes are included in the radiation transfer 
 code {\sc cloudy}.

\subsection{Thermal equilibrium and the evolution timescale}

\begin{figure}
\includegraphics[scale=0.45]{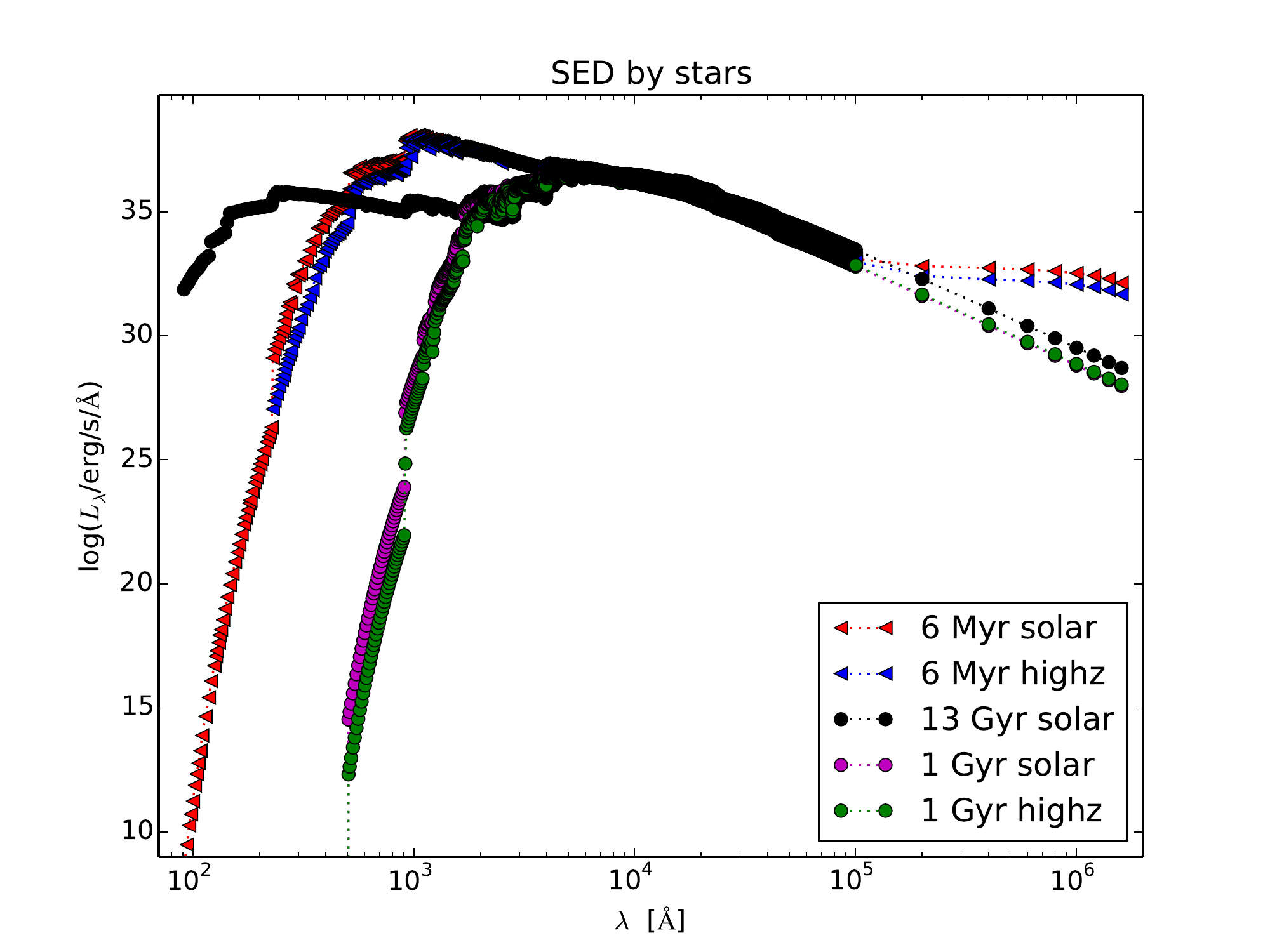} 
\caption{The broad band continua of the NSC stellar-energy
 distributions used in our computations. Red triangles represent the SED from the 6~Myr stellar population embedded within the NSC around 
Sgr~A*, while black dots represent the SED for the 13 Gyr stellar population of M60-UCD1.
This radiation component contributes to the cooling rate of the ISM compared to the mechanical 
heating by winds from the hot stars and by radiation originating  from the central source of 
the SMBH accretion.
In addition, the SED from a possible older, 1~Gyr, stellar population is presented by magenta dots, but the total luminosity of this cluster is two orders of magnitude lower and has no effect on TI of Sgr~A*. For 6~Myr and 1~Gyr cases, solar versus increased metallicity are compared. 
An increase in 
metallicity of up to z=0.04 does not change the shape of the continuum significantly.}
\label{fig:spec1}
\end{figure}

Depending on the profile of the spectral energy distribution (SED), the incident radiation field can
 both cool and heat the plasma. The net effect is determined by plasma 
density as well as the spectral shape of the incident 
radiation. Thus, the presence of the radiation field contributes both to 
 the cooling and heating rates, i.e., ${\cal L}$ and ${\cal H}$, 
respectively   in erg s$^{-1}$ cm$^{-3}$. 
The difference between these two parameters describes the rate of exchange of total energy 
per unit volume  by different cooling and heating processes.

We include the stellar emission from the NSC by modeling its spectral shape appropriate for a given age.
We also include the radiation produced by the accretion flow onto the central black hole, which is 
highly concentrated towards small radii. 
Nevertheless, our model is still an idealized scenario. 

\begin{figure*}
\includegraphics[scale=0.45]{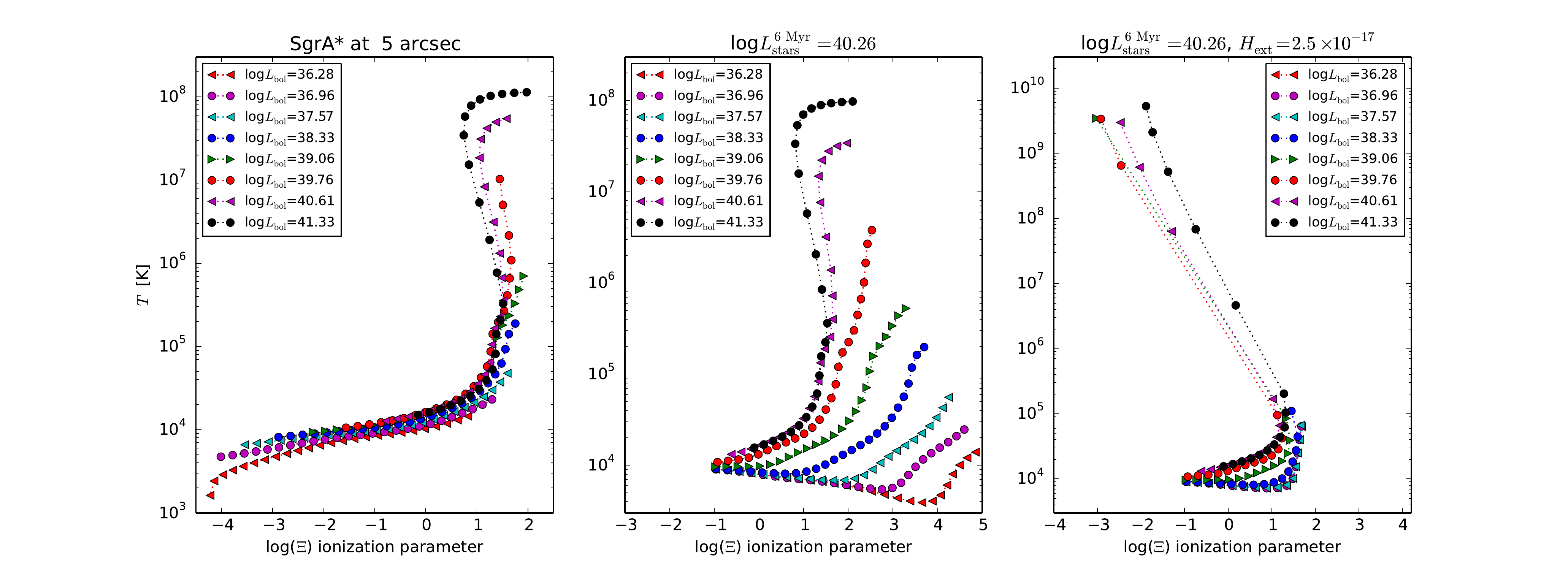} 
\includegraphics[scale=0.45]{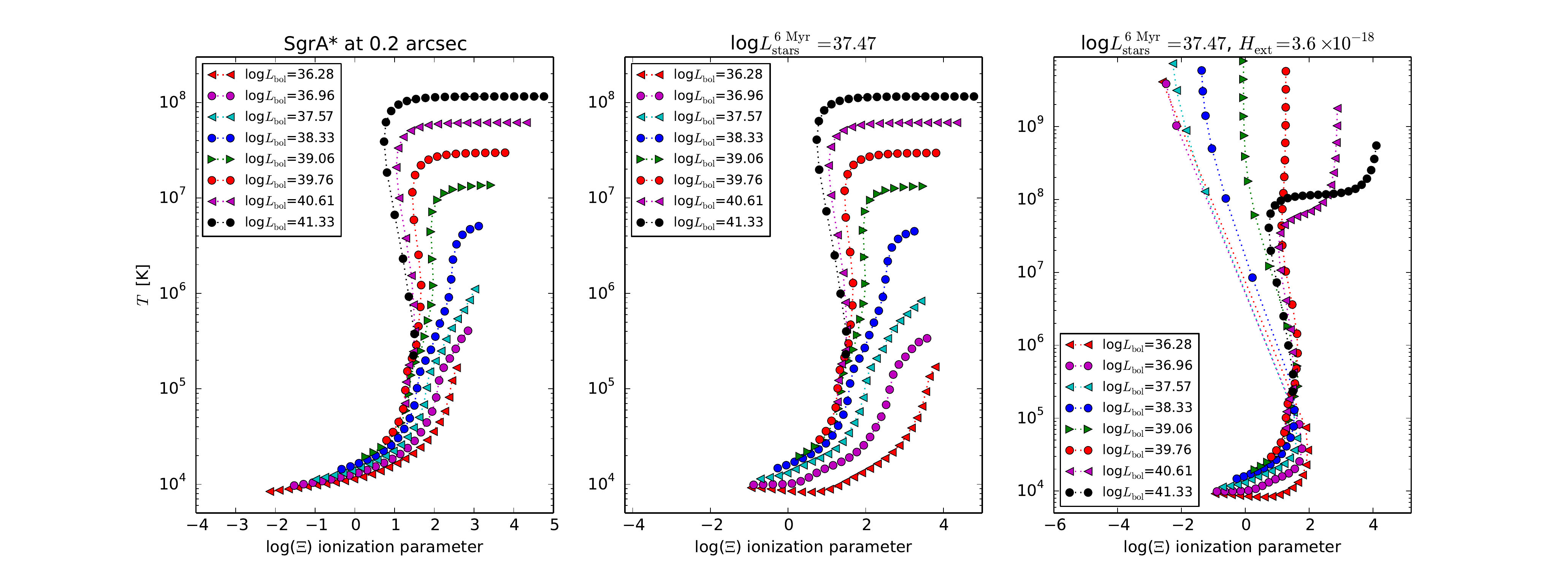} 
\caption{Solutions for S-curve of TI in the plane of temperature vs. ionization parameter, as defined 
in Eq.~\ref{eq:bigxi}, for different luminosity states of the radiation: from the central source  only (left panels), 
together with heating by stellar radiation (middle panels), and together with mechanical heating 
by winds (right panels). Values of central source luminosity are marked within the panels.  
We present results for the gas located at 5~arcsec from  Sgr~A* (upper row of panels) and at 
0.2~arcsec (bottom row of panels).  The luminosity and volume mechanical heating are 
log($L_{\rm stars}$/erg~s$^{-1})=40.03$ and $H_{\rm ext}=2.5 \times 10^{-17}$ erg s$^{-1}$ cm$^{-3}$, respectively, at 5~arcsec 
and log($L_{\rm stars}$/erg~s$^{-1})=37.47$ and $H_{\rm ext}=3.6 \times 10^{-18}$ erg s$^{-1}$ cm$^{-3}$,  respectively, 
at 0.2~arcsec from Sgr~A*.}
\label{fig:SgrA*_equilibrium}
\end{figure*}

 Similar processes of mass accumulation and generation of outflows 
with effects of stellar feedback have been explored on larger scales, typically from 
length-scales of several tens to several $\times10^2$~pc \citep{2009MNRAS.393..759S}, but in 
this paper we are interested only in the innermost regions of the NSC (typically a fraction of a parsec). 
For a given source our approach allows us to adjust the 
mechanical heating rate ${\cal H}$  according to observational estimates. 

We find the thermal equilibrium solution from the balance between all cooling and heating
terms,  
\begin{equation}
\dot{Q} = {\cal L} - {\cal H} = 0.
\label{dotq}
\end{equation}
In a stationary system, the medium is thermally stable if the following condition for entropy perturbation 
$\partial s$ is fulfilled \citep[][chapt.~7]{dopita2003}:
\begin{equation}
\frac{\partial (\dot{Q}/T)}{\partial s} >0,
\label{eq:ti1}
\end{equation}
where $\dot{Q}$ combines the heating and cooling mechanisms operating in the given system. For
 isobaric perturbations, Eq.~(\ref{eq:ti1}) reduces to
\begin{equation}
\left.\frac{\partial\dot{Q}}{\partial T}\right|_{P}>0,
\end{equation}
which is equivalent to the original \citet{field1965} criterion, assuming the perfect gas equation 
of state.

To solve for the stability of the solution to Eq.~(\ref{dotq}), we employ {\sc cloudy} 13.02 
\citep{ferland2013}, where the mechanical volume heating is included through 
the appropriate option {\it Hextra}
 (see Hazy1 and Hazy2 documentation files), which reflects all possible
 heating of gas due to stellar winds of stars located within the region of consideration\footnote{http://www.nublado.org/}. 
This solution provides us with the local equilibrium temperature of the medium as a function of 
assumed ISM hydrogen density at a given location. The stability can be conveniently estimated from 
the diagram of stability curve. Note that in the
definition of the ionization parameter $\Xi$ \citep{tarter69},  we need to also include 
the luminosity of the stellar radiation field, which accounts for the gas ionization state: 
\begin{equation}
\Xi = { L_\bullet + L_{\rm stars } \over  4 \pi c P_{\rm gas} R^2},
\label{eq:bigxi}
\end{equation}
where $L_\bullet$ is the accretion luminosity on to SMBH, $L_{\rm stars}$ is the stellar luminosity, 
$P_{\rm gas}$ is the local gas pressure, and $R$ is the distance from the
center. 
The curve $\Xi$ vs.\ $T$ indicates the thermal stability: in this relation, the branch with 
the positive slope is radiatively stable while the branch with the negative slope is unstable.

The cooling timescale to reach equilibrium (or time for a significant departure 
from it) is estimated as
\begin{equation}
t_{\rm cooling} = {E \over  \cal L}.
\label{eq:cool}
\end{equation} 
In fact, this expression underestimates the actual timescales when the system is very close to 
thermal equilibrium, i.e.\ when ${\cal H}\approx{\cal L}$.
 Nonetheless, for a broad range of parameters,  the cooling timescale is a reliable estimate 
 of the rate at which TI develops in the system. 

\subsection {Non-equilibrium solutions}

In general,  the entropy $s$ of matter evolves according to the equation
\begin{equation}
\frac{ds}{dt}=\frac{{\cal L}-{\cal H}}{n T}\propto \frac{1}{t_{\rm cooling}}.
\label{eq:entropy}
\end{equation}
 The corresponding timescale for cooling,  
$t_{\rm cooling}$, now depends on the difference between the cooling and 
heating rates, $\cal L-H$, 
and has to be calculated for initial conditions of the plasma far from thermal equilibrium.
To streamline the analysis, the cooling time-scale is frequently expressed as a function of 
temperature only.

For very short cooling times the initial perturbations condense rapidly into cold clumps, which then 
gradually fall down toward  the centre in the gravitational field of the SMBH. In the opposite 
direction, bubbles of gas rise outwards in the hot phase.
This is the same mechanism as the one operating in the context of intergalactic gas in galaxy 
clusters \citep{mccourt2012}. If the timescale to establish equilibrium greatly exceeds the 
dynamical timescale of accretion and/or outflow, 
named the crossing timescale $t_{\rm cross}$, the supplied material cannot reach equilibrium temperature. 

\section{Results}
\label{sec:res}
\subsection{Case of Sgr~A*}

\begin{figure*}
\includegraphics[scale=0.45]{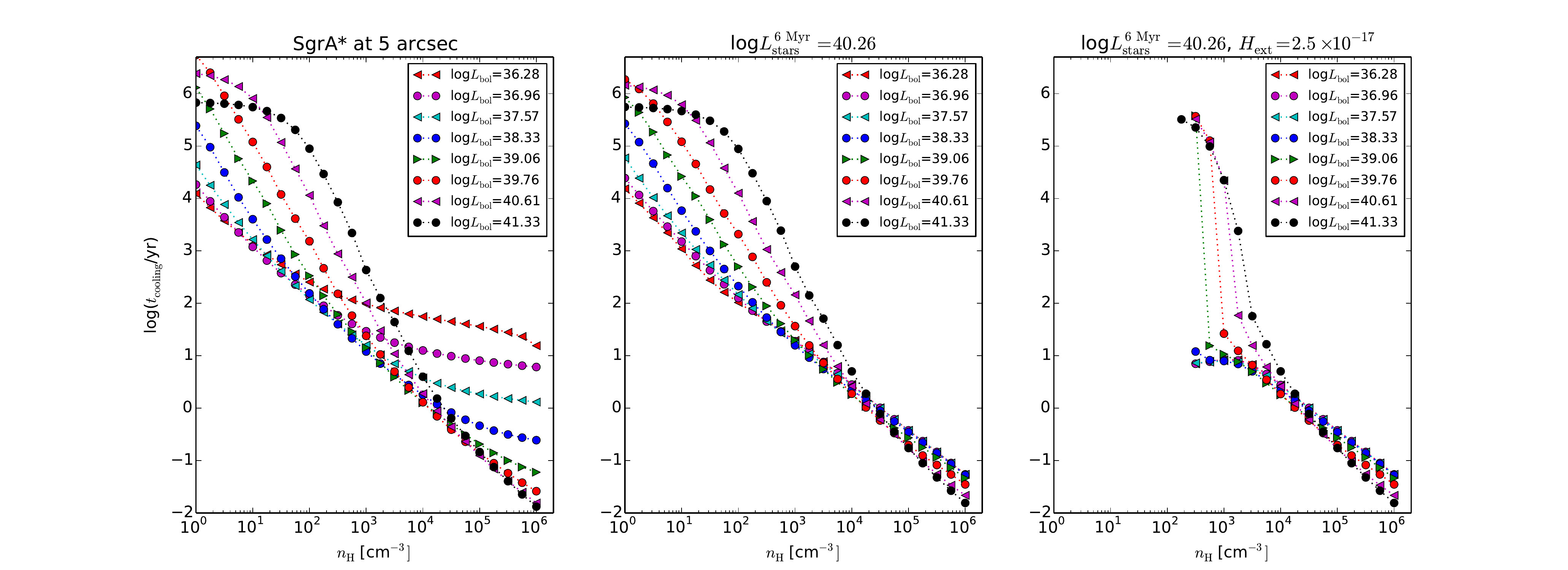} 
\includegraphics[scale=0.45]{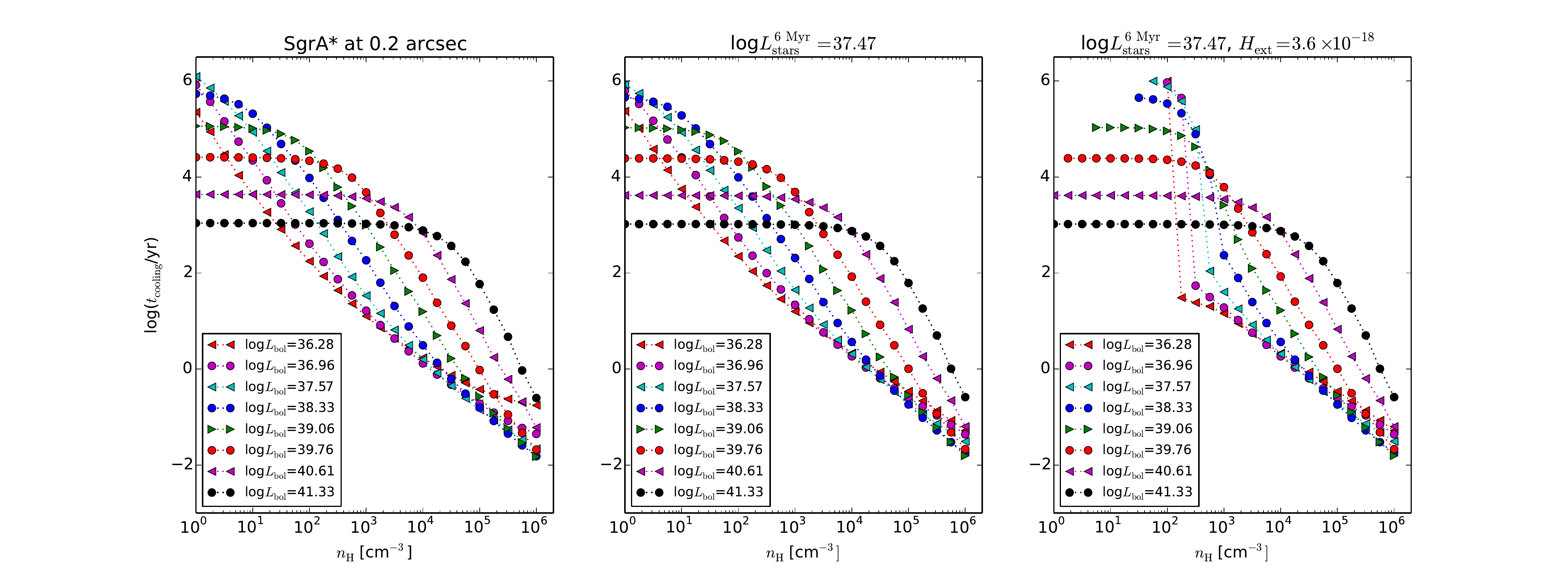} 
\caption{Equilibrium cooling timescale in the case of Sgr~A* 
for different luminosity states of the radiation: from the central source  only (left panels), 
together with heating by stellar radiation (middle panels), and together with mechanical heating 
by winds (right panels).  The results presented here are for the same model parameters as in Fig.~\ref{fig:SgrA*_equilibrium}.}
\label{fig:SgrA*coooling}
\end{figure*}

We assess the possibility of the existence of a multi-phase equilibrium taking into account 
three effects: (i)~radiative heating by the accreting central black hole, as in \citet{rozanska14}; 
 (ii)~additional radiative heating by stars;   
(iii)~and  mechanical heating by stellar  winds. 

Since Sgr~A* was likely 
in much higher luminosity states in the past several hundred years 
\citep[see][and references therein]{ponti2010}, we consider eight possible bolometric
 luminosities   (always listed in the plots)
 of the central source, from the lowest one  in quiescence: log$(L_{\rm bol})= 36.28$,
 to the highest in the active phase: log$(L_{\rm bol})=41.33$ \citep{zhang2015a,czerny2013c}. 
The spectral shape of the incident central emission depends on the bolometric luminosity. The final shape of the spectrum is computed from the theoretical model of radiatively inefficient Bondi
accretion flow for eight values of the accretion rate.
 Spectra used for our photoionization  calculations are the same as presented in Fig. 5 of \citet{moscibrodzka2012} .

We describe the stellar light using the 
spectral template for stars from {\sc Starburst99} 
\citep[][see Fig.~\ref{fig:spec1}]{leitherer1999,lu2013}. 
As our basic model, we consider a young stellar 
population of 6~Myr stars with solar metallicity. We also 
checked the effect of higher metallicity stars and of the presence of an additional older stellar 
population of 1~Gyr stars. 
The normalization is set by the assumption that  the total mass of the young NSC is $\sim 3.7\times 10^4$~M$_{\odot}$ within 0.5~pc \citep{lu2013}, whereas the total mass of old NSC is $\sim 8.0 \times 10^6$~M$_{\odot}$ within 2.5~pc \citep{chatz2015}. 
The canonical spectral shape of stellar radiation used as an 
input to our  {\sc cloudy} simulations of the gas conditions around Sgr~A*
is presented in Fig.~\ref{fig:spec1} by red triangles.

{\sc cloudy} code always takes the illuminated source as a point source but the star cluster is an extended source, and the fraction of 
the ionization parameter, which is computed from the luminosity of stars is, constant over the whole NSC volume. 
Therefore, to ensure that the {\sc cloudy} takes the same ionization parameter for all radii we within the nuclear cluster (which is 
physically consistent),  we have to renormalise stellar luminosity depending on the distance, which 
is also an input parameter for {\sc cloudy} simulations. In practice, we take the following steps:
i)  compute the SED for a given population and total mass of the NSC (given above) using Starburst99, ii) integrate this SED to obtain  the total luminosity from all stars - $L_{\rm stars}$, iii) assume that everywhere  within the NSC gas "sees" the 
same amount of radiation, so the ionization parameter: $\Xi = L_{\rm stars } /( 4 \pi c P_{\rm gas} R^2)$ is the same,
iv)  by changing input parameters for {\sc cloudy}, which means that to keep  the ionization parameter the same, and for the same density, 
changing the distance (for SgrA*, from 0.2 arcsec to 5 arcsec) requires changing the input $L_{\rm stars}$. 
The appropriate values of stellar luminosities are always listed 
in the figures.

The ``mechanical'' heating by stars is described following the approach of \citet{shcherbakov2010}.
As in their Fig.~3, we adopted values for the stellar mass input into the feeding region
around BH equal to 
$10^{-8} M_{\odot} {\rm yr}^{-1}$ and wind velocity $1000$~km~s$^{-1}$ at a distance of 
0.2~arcsec from the nucleus.   The mass input at a distance of 5~arcsec is 
$3 \times 10^{-5} M_{\odot} {\rm yr}^{-1}$, and the wind velocity equals  $1200$~km~s$^{-1}$.
This leads to a mechanical heat input of
\begin{equation}
H_{\rm ext} = \left\{\begin{array}{ll} 
3.6\times10^{-18}\, [{\rm erg\,s}^{-1} {\rm cm}^{-3}] \qquad {\rm at} &\; r = 0.2~{\rm arcsec}, \\
\\
 2.5\times10^{-17}\, [{\rm erg\,s}^{-1} {\rm cm}^{-3}] \qquad {\rm at} & \; r = 5~{\rm arcsec}.  
\end{array}\right.
\end{equation} 

 The description of the heating in the inner region is a crude approximation, since 
it relies on a single star being present in this region. As in \citet{shcherbakov2010}, 
we neglect here the velocity of the star itself,
 as the fate of the lost material is an open issue.
 The velocity of this star is likely higher
(the Keplerian velocity at a radius of 0.008 pc is 1500 km s$^{-1}$ for a $4 \times 10^6 M_{\odot}$) 
than the wind velocity  (assumed to be 1000 km s$^{-1}$ here). Such a wind is partially focused in the 
direction of the stellar
motion. So on one hand we should assume that the effective velocity of the wind is enhanced due to the 
stellar velocity, but on the other hand part of the material may leave the region as a stream and not 
dissipate all the energy in situ. Since the wind mass loss is provided as an order of magnitude estimate, 
we do not include the stellar motion into the heating term.


As mentioned above, we study the thermal equilibrium conditions around Sgr~A* at two representative radii, 5.0 and 
0.2~arcsec, corresponding to a linear distance of about 0.2~pc and 0.008~pc from the center, respectively. 
The stability curve is obtained by running a grid of {\sc cloudy} simulations for constant 
hydrogen number density clouds spanning a range from $\log n_{\rm H}=0$ up to $\log n_{\rm H}=6.25$.
The results for both distances are shown in Fig.~\ref{fig:SgrA*_equilibrium} (upper and lower 
panels respectively). The left panels show the results when only radiative heating by the 
accreting  material affects the ionization equilibrium. The present  level activity
of Sgr~A*  is too low to support a high temperature of the medium, and only the cold  equilibrium branch exists. 
A higher level of activity in the past, however, could have heated the plasma up to the Inverse 
Compton temperature, i.e. $T \sim 10^8$~K.

The addition of stellar light (middle panels in Fig.~\ref{fig:SgrA*_equilibrium}) to radiative 
heating by the central source results in a decrease of temperature of the irradiated medium.
 Stellar light thus effectively cools the gas even though the stars are blue and young. Since the 
 central emission has a strong X-ray component, the addition of a blue light provides additional 
soft photons, which lowers the Inverse Compton temperature of the incident radiation.

 We have checked that higher metallicities in the stellar population does not change 
our results. The SED of high metallicity stars  is not very different from the solar case 
(see Fig.~\ref{fig:spec1}), and the resulting plots for the S-curves did not show any noticeable 
differences in comparison to Fig.~\ref{fig:SgrA*_equilibrium}. Also the contribution of an older stellar 
population of 1~Gyr stars does not change quantitatively the location of TI presented in 
Fig.~\ref{fig:SgrA*_equilibrium} since those stars are two orders of magnitude less luminous (despite their larger total mass).
But the inclusion of mechanical heating by stellar winds changes the picture 
dramatically. We demonstrate this in the right panels of Fig.~\ref{fig:SgrA*_equilibrium}.
Although we were not able to cover the entire low-density range for the case of high-luminosity 
and large distance, even for the displayed  span of density the equilibrium temperatures  exhibit 
unrealistically extreme values.  
 Cold clumps can exist in  thermal equilibrium but they must be very dense. Closer to the black hole 
the stellar mechanical heating is one order of magnitude lower, the radiative effects are 
of greater importance, and we still have the upper stable branch of solution at the 
Inverse Compton temperature for the highest luminosity states.
Again, an increase in metallicity of the stellar population or the inclusion of an additional older population of stars, do not change our results.

The present-day activity level  of Sgr~A*  supports temperatures up to $10^5$~K on the stable branch
but only for densities higher than $n_{\rm H} \sim 10^2$ cm$^{-3}$
\citep[e.g.,][and further references therein]{wang2013,rozanska2015}. 
Lower-density gas gets heated to a much higher temperature even at a small distance from
 the black hole. The time taken for the low-temperature phase to reach equilibrium is long.
 In Fig.~\ref{fig:SgrA*coooling} we show the cooling timescales for plasma in equilibrium:  
 for densities lower than $10^2$ cm$^{-3}$ the cooling timescales exceed $10^4 $~yr.

This has to be compared with the characteristic timescale of inflow and outflow. The stagnation radius, 
dividing the inflow from outflow in Sgr~A*, is currently located between 0.3 to 1~arcsec
 \citep{quataert2004,shcherbakov2010}.   Thus our solution at 0.2 arcsec is within the accretion flow
part. The inflow proceeds initially with a velocity which is a small fraction of the stellar wind 
velocity (zero speed formally at the stagnation radius). 
Assuming that the inflow velocity is 10\% of the of the stellar wind velocity, 
i.e. $\sim 100$ km/s, we obtain an 
infall timescale of 100~yr,  
much shorter than the cooling timescale. Thus, close to the nucleus of Sgr~A*, 
the adiabatic picture of the hot plasma as presented by \citet{quataert2004} is satisfactory. However, if cold, sufficiently 
dense clumps of gas happen to arrive in that region from outside they will also survive due to the presence of the cold equilibrium branch.  

The solution at 5 arcsec is in the outflow part, above the stagnation radius. The outflow 
proceeds predominantly with a speed comparable to the wind speed. Again, we see that there is no 
time for a low density plasma to cool down  unless the outflow finally  stops due to collision 
with the outer ambient medium that is initially far from thermal equilibrium. 

Stars supply a moderately hot, $T \sim 10^6$~K, low-density material.
This temperature does not coincide with the equilibrium solution 
(see Fig.~\ref{fig:SgrA*_equilibrium}) and the plasma may remain away from the 
thermal equilibrium curve since the timescales for reaching thermal balance are long 
(see Fig.~\ref{fig:SgrA*coooling}). We thus address separately the  issue of 
thermal evolution of this freshly injected plasma.
 
\begin{figure}
\includegraphics[scale=0.51]{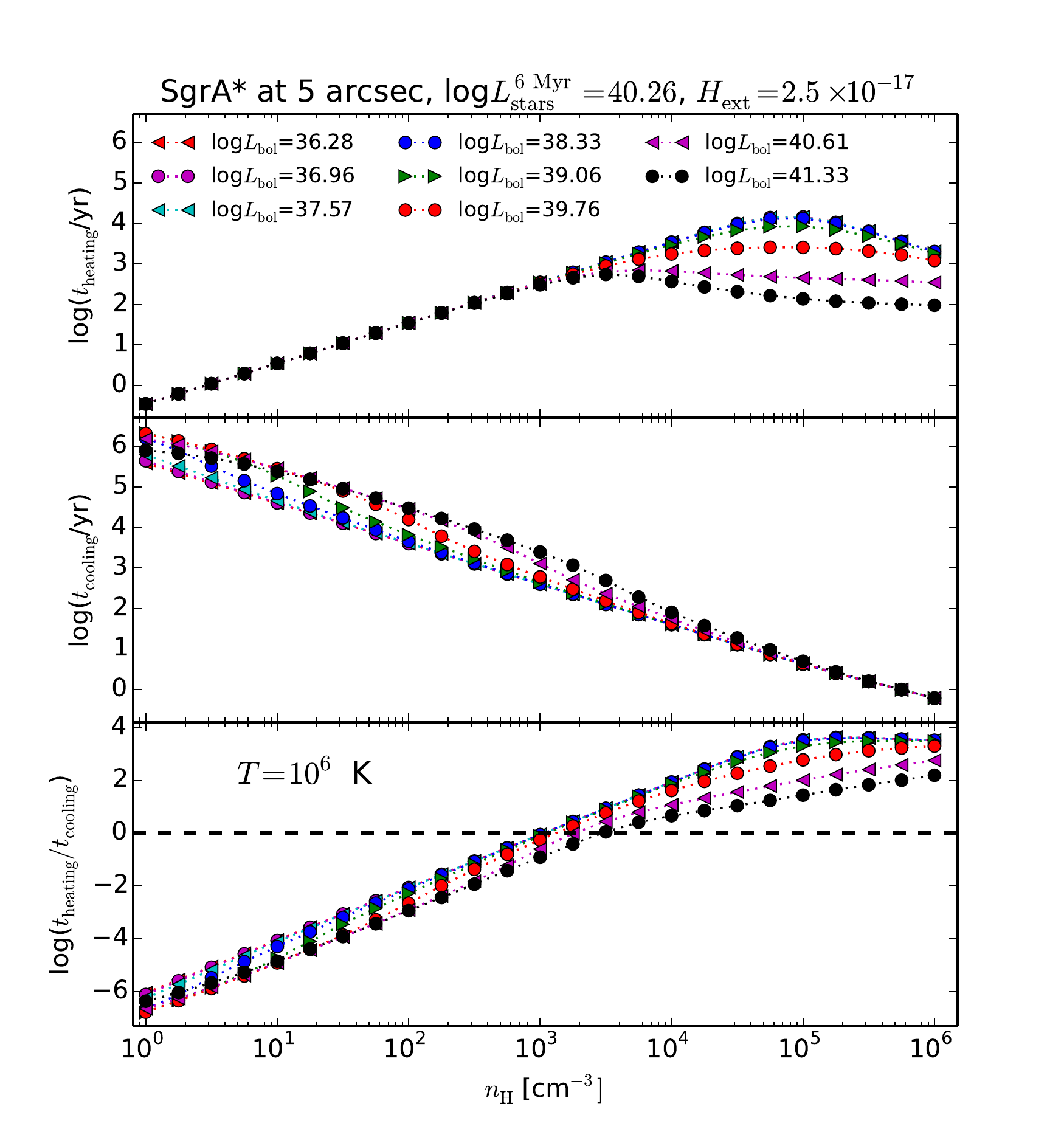} 
\caption{Sgr~A* heating (upper panel) and cooling (middle panel) time-scales plotted against cloud 
density. Bottom panel represents ratio of those two values. The ISM gas in each case is 
heated by radiation from the center with different luminosity states, radiation from stars 
log($L_{\rm stars}$/erg~s$^{-1}) = 40.03$, and volume mechanical heating
 $H_{\rm ext} = 2.5 \times 10^{-17}$ erg s$^{-1}$ cm$^{-3}$.
Each cloud is assumed to have a temperature of $T=10^6$ K and is located at} an offset of
5~arcsec from the center. 
\label{fig:SgrA*coooling_noneq}
\end{figure}

\begin{figure}
\includegraphics[scale=0.51]{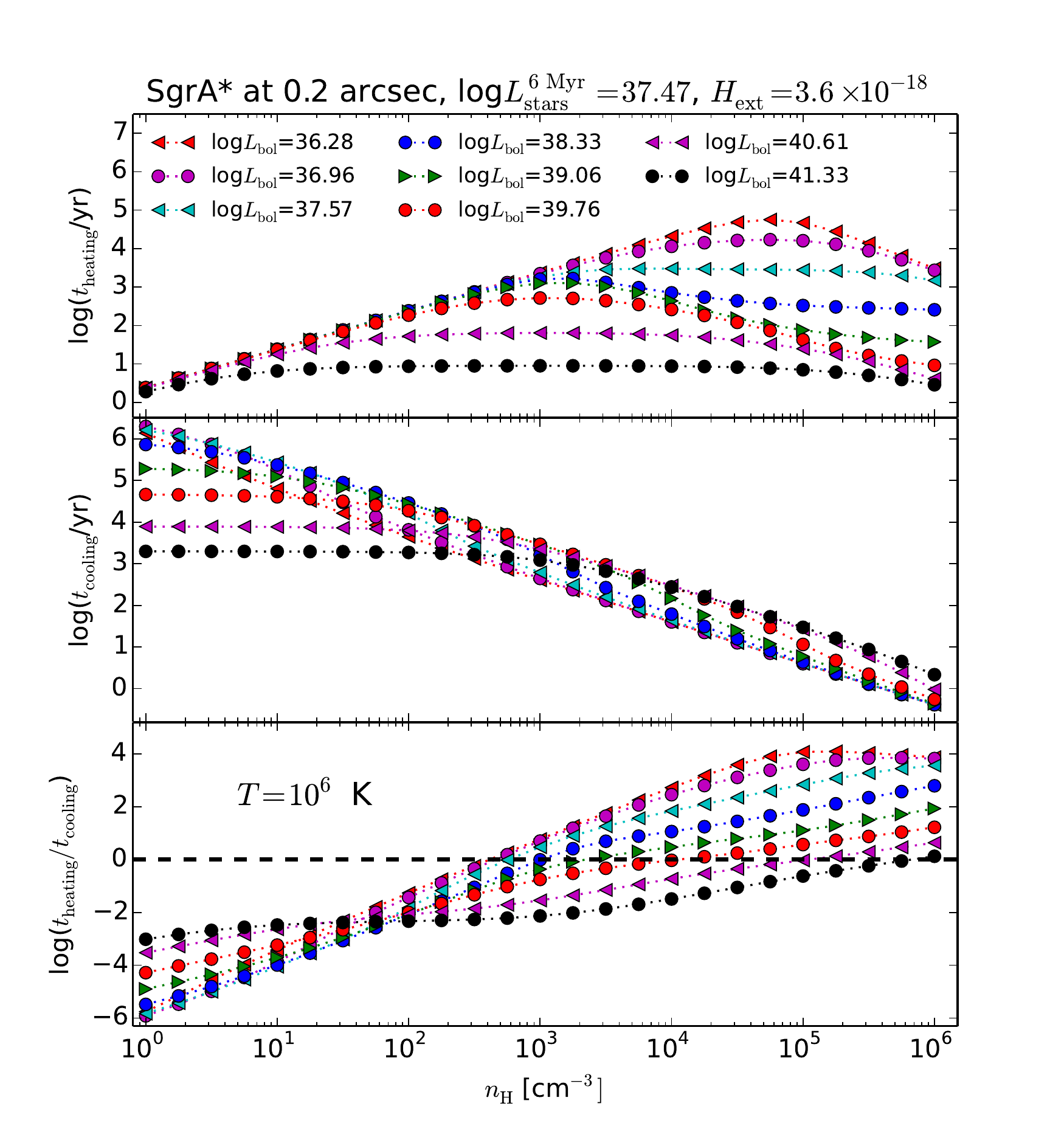} 
\caption{Sgr~A* heating (upper panel) and cooling (middle panel) time-scales plotted against cloud 
density. Bottom panel represents ratio of those two values.  The ISM gas in each case
is heated by radiation from the center with different luminosity states,  radiation from stars 
log$(L_{\rm stars}) = 37.47$ erg s$^{-1}$, and volume mechanical heating
 $H_{\rm ext} = 3.6 \times 10^{-18}$  erg s$^{-1}$ cm$^{-3}$. Each cloud is assumed to have 
a temperature of $T=10^6$ K and is located at a distance of 0.2~arcsec from the center. }
\label{fig:sgrthermal02}
\end{figure}

For this, we fix the plasma
temperature at $10^6$ K and  calculate the cooling timescale as a function of density, at both 
representative radii, 5 and 0.2 arcsec. The results are shown in 
Figs.~\ref{fig:SgrA*coooling_noneq} and \ref{fig:sgrthermal02}, respectively. 
The upper panels show timescales computed with Eq.~(\ref{eq:cool}) assuming the total heating given 
as an output in {\sc cloudy} computations, middle panels show the same for cooling only, and
the bottom panels show ratios of those two values.  The cooling timescales for a 
very dense plasma are  relatively short in comparison to the heating timescales. Therefore, we 
conclude that for densities higher than $\sim 10^3$ cm$^{-3}$, cooling dominates over heating.
Recent X-ray observations of hot plasma in the vicinity of Sgr~A* indicate 
that the plasma density is of the order of 10--100 cm$^{-3}$ \citep{shcherbakov2010,rozanska2015}, 
and for such low densities heating dominates.
The timescales are short, of  the order of 100 years,
and in the case of the smaller radii  (Fig.~\ref{fig:sgrthermal02}), they significantly depend on the 
actual luminosity state of Sgr~A*. Therefore, vigorous hot plasma outflow is expected from most of the region, both at the present level
of moderate Sgr~A* activity and in the past.
Occasional strong wind collisions may result in some local plasma 
compression and formation of clouds but the requested compression factor 
is large, so the phenomenon 
is unlikely to be frequent.

\begin{figure}
\includegraphics[scale=0.51]{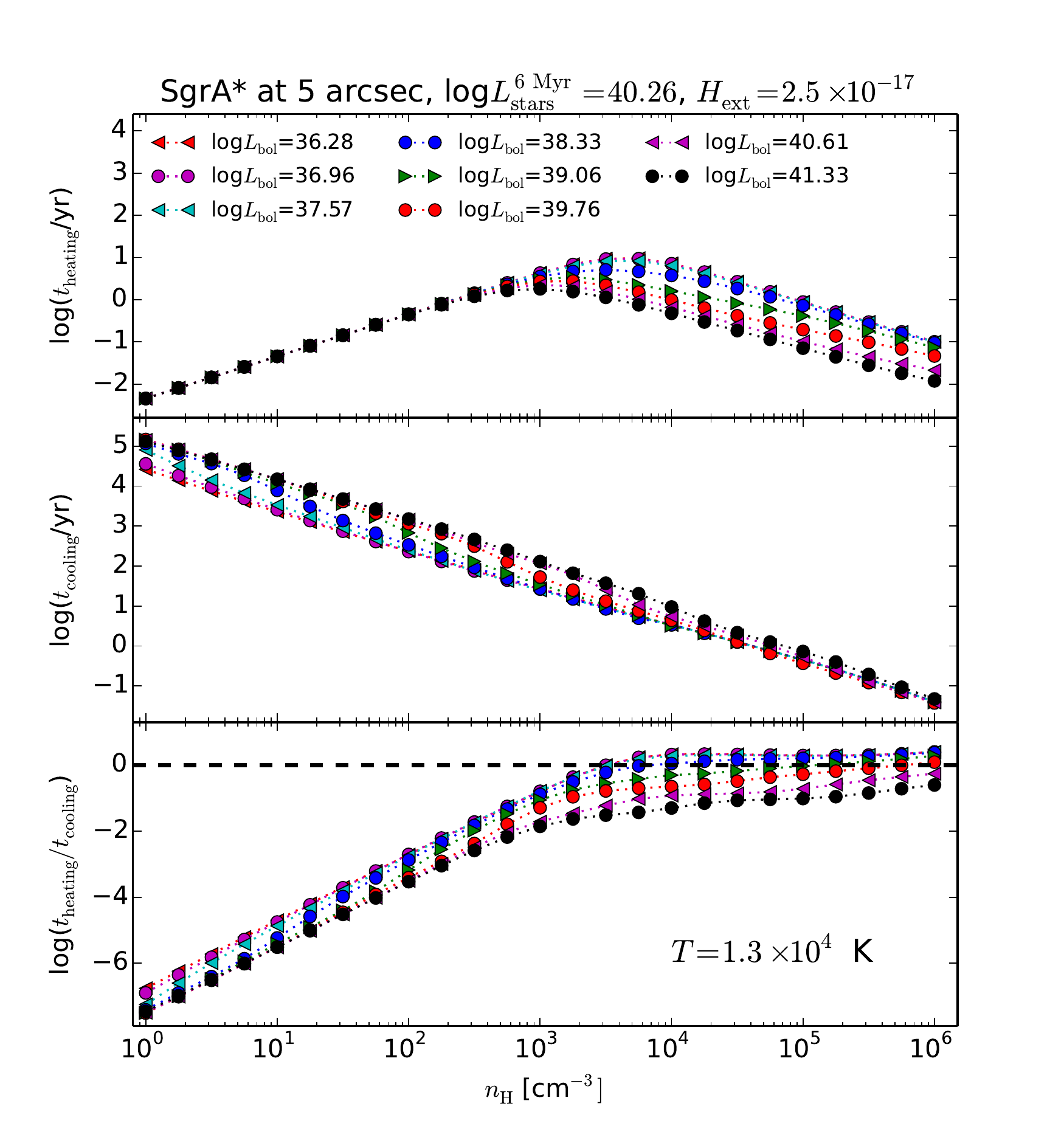} 
\caption{The same as in Fig.~\ref{fig:SgrA*coooling_noneq} but each  cloud is assumed to have 
temperature $T=1.3 \times 10^4$ K.}
\label{fig:SgrA*coooling_noneq_4}
\end{figure}

\begin{figure}
\includegraphics[scale=0.51]{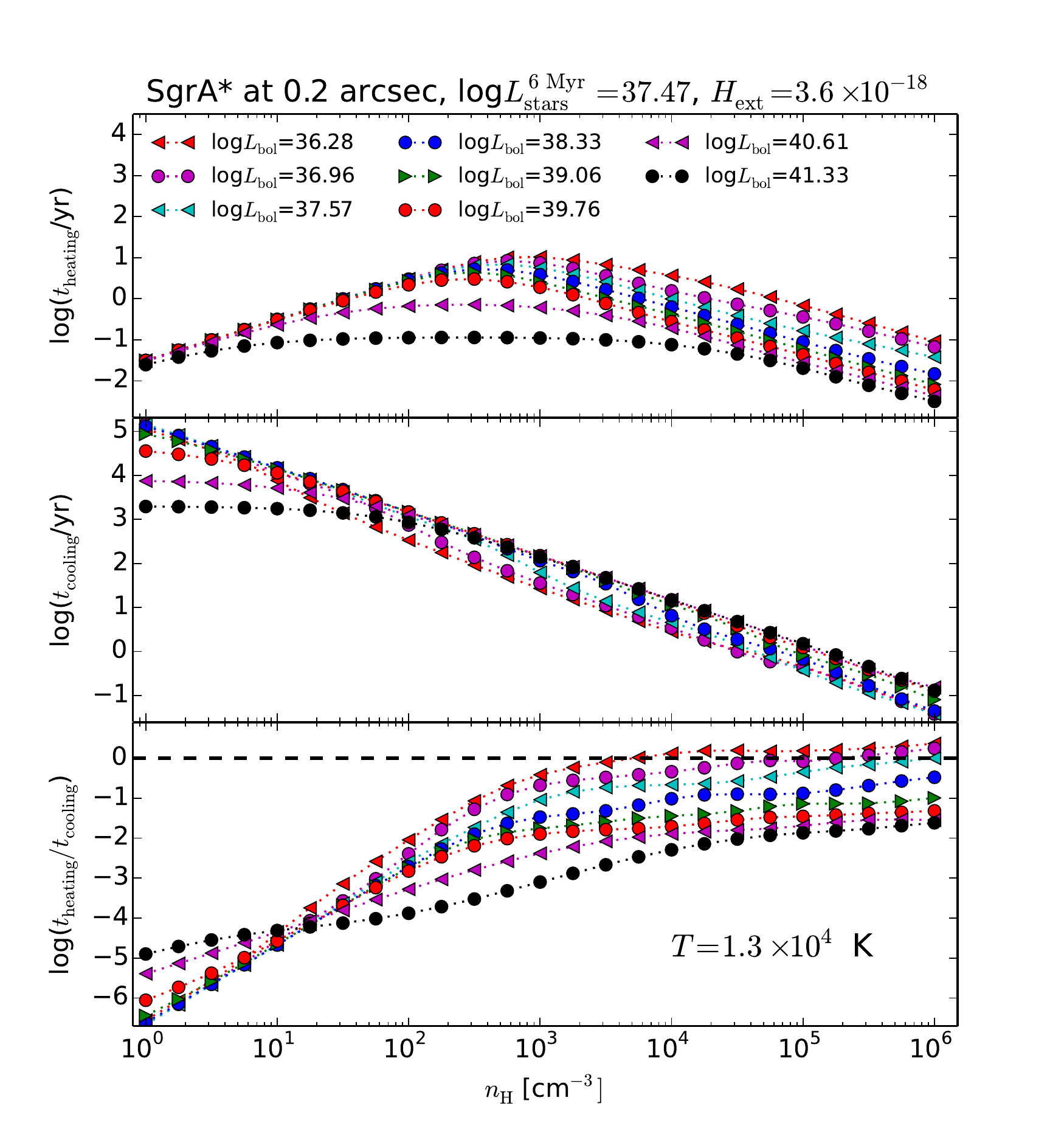} 
\caption{The same as in Fig.~\ref{fig:sgrthermal02} but each  cloud is assumed to have 
temperature $T=1.3 \times 10^4$ K.}
\label{fig:sgr4thermal02}
\end{figure}

 Spontaneous formation of cold clouds from the low density medium is not expected.  On the other 
 hand, in the GC mini-spiral region numerous cold dense clouds do exist, as an intense IR 
 emission from that region is observed  \citep{becklin1978,gezari85}. 
 We thus repeated a similar non-equilibrium
  study as above, but for clouds with temperatures of  
  $~1.4 \times 10^4$ K, determined observationally from the mini-spiral. As in the previous two figures, heating and cooling timescales for matter located at 5 and 0.2~arcsec from the center are presented in 
Figs.~\ref{fig:SgrA*coooling_noneq_4} and \ref{fig:sgr4thermal02} respectively. We see that clouds
 denser than $10^4$ cm$^{-3}$ reach equilibrium f in timescales shorter than 100 years and preserve 
 it. Such clouds are thus located  in the lower part of the equilibrium curve.  
 They will be surrounded by a rapidly expanding/outflowing hot phase. 
  The interaction of colder clouds with 
 the hot medium may finally lead to cloud destruction but this issue is far beyond the scope of 
 this paper, as the phenomenon will likely depend on the actual cloud motion, and likely also on the 
 magnetic field intensity. Computations presented in this paper only show that the cold medium phase of the 
 mini-spiral is not produced 'in situ' and had to form in a different way than  due to thermal 
 instability in the hot plasma heated by radiation and stellar winds.

\subsection{Case of M60-UCD1}


\begin{figure}
\includegraphics[scale=0.45]{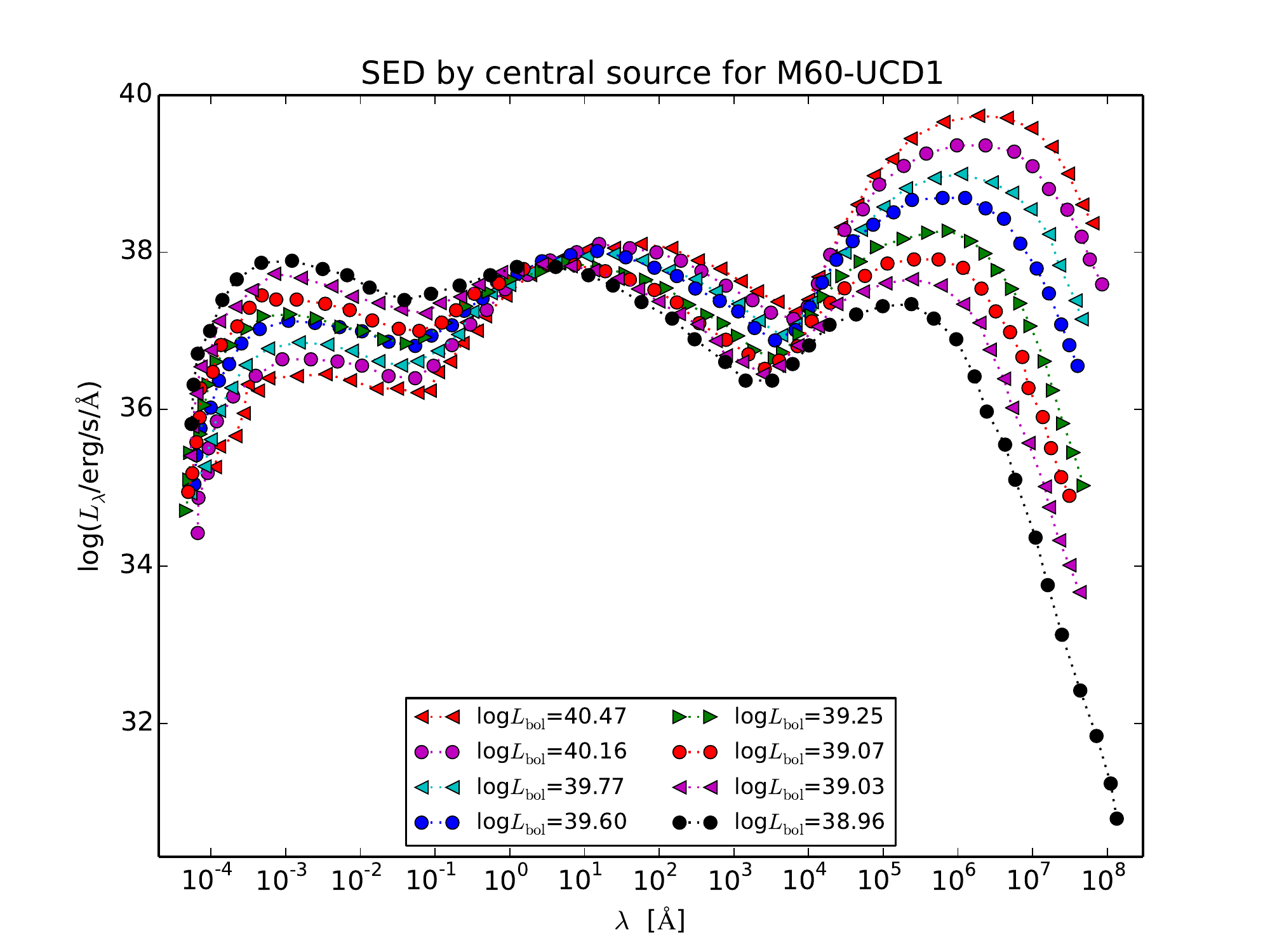}
\caption{The M60-UCD1 broad-band continuum assumed in the computations.
The shape of different luminosity states for Sgr~A* is normalized  to take into account 
the observational value of the X-ray flux for M60-UCD1: $L_{\rm X}$(2-10 keV) = $1.3 \times 10^{38}$
erg/s. }
\label{fig:spec2}
\end{figure}

Ultra-Compact Dwarf galaxies, apart from the characteristic compact size of their dense nuclear 
stellar cluster generally exhibit elevated mass-to-light ratios \citep{fellhauer06,mieske2014}. 
M60-UCD1 is an exceptional example with tentative evidence, based on stellar kinematics, 
for a supermassive black hole in the core \citep{strader2013,seth2014,reines14}.
The object has also been detected as a variable X-ray source \citep{Luo2013}. 
We can thus apply to this source a similar  analysis as in the previous section. 

In the first step we again construct a model of the broad-band spectrum emitted from  
the accretion inflow from \citet{moscibrodzka2012}. But in case of M60-UCD1, 
the X-ray luminosity between 2--10 keV is $1.3 \times 10^{38}$ erg s$^{-1}$ \citep{strader2013}, i.e.,
 higher than in Sgr~A*. Therefore,  we re-scale the broad-band spectra used in  the Sgr~A* calculations 
 to obtain the observed value of the X-ray luminosity between 2-10 keV for this source.
 All eight different luminosity states used in our photoionization calculations are presented 
 in  Fig.~\ref{fig:spec2}. In the second step we model the stellar emission of
 M60-UCD1 nuclear star cluster with {\sc Starlight} \citep{fernandes2011}, 
 taking the total mass of the stars to be 
  $1.2 \times 10^8 M_{\odot}$  \citep{seth2014} and assuming the age of the stars to be 13~Gyr 
  \citep{strader2013}. The SED from stars used in {\sc cloudy} computations is presented in 
Fig.~\ref{fig:spec1} as black dots.

\begin{figure*}
\includegraphics[scale=0.45]{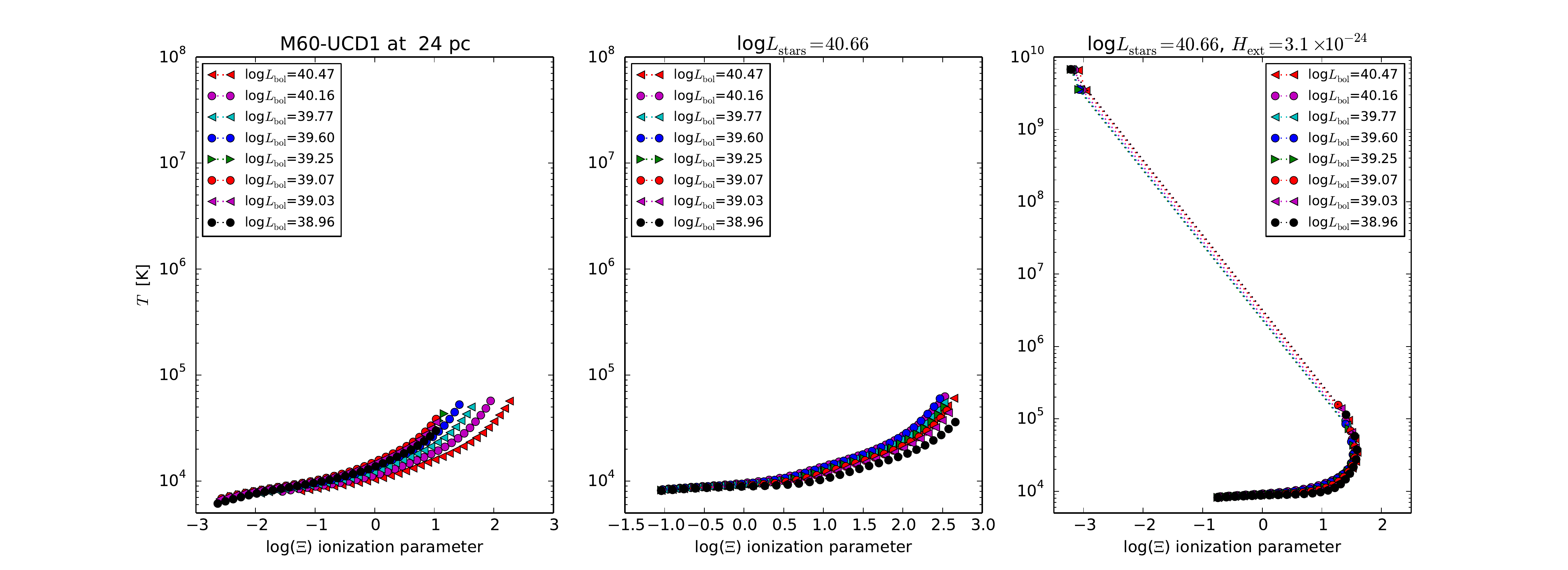} 
\includegraphics[scale=0.45]{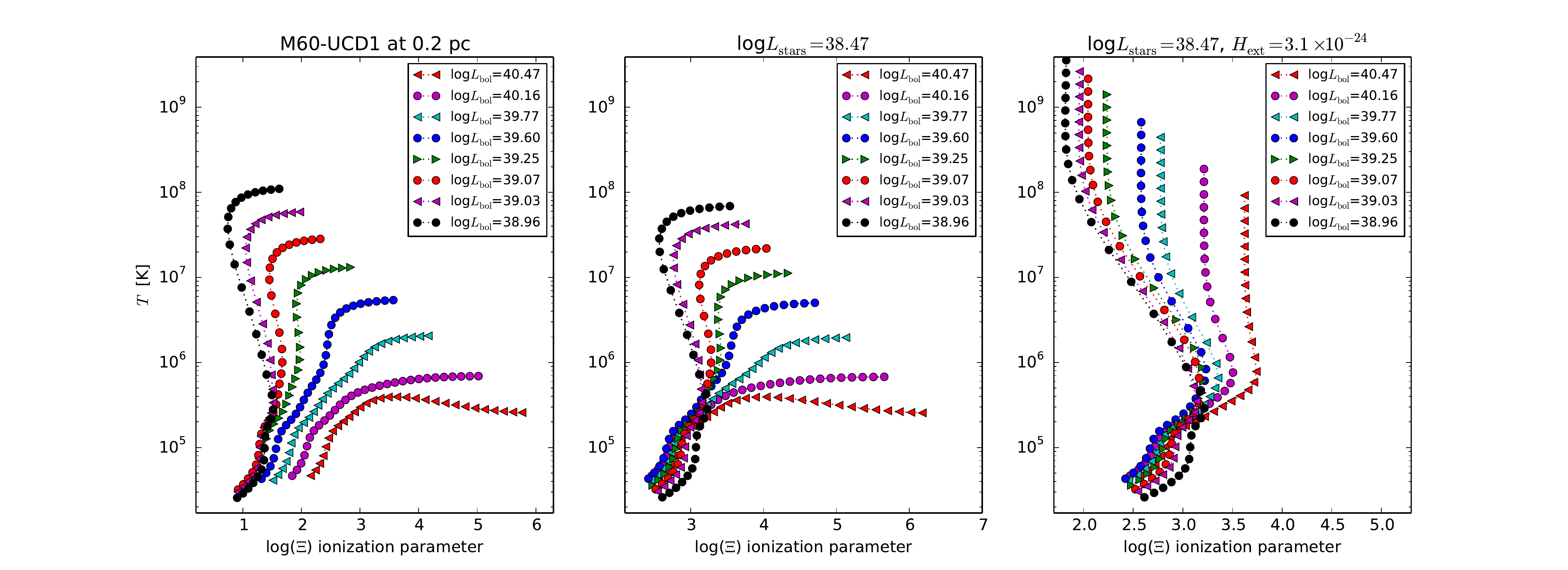} 
\caption{Solutions for S-curve of TI in the plane of temperature vs. ionization parameter, as defined 
in Eq.~\ref{eq:bigxi}, in case of  M60-UCD1, for different luminosity states of the radiation:
 from the central source  only (left panels), 
together with heating by stellar radiation (middle panels), and together with mechanical heating 
by winds (right panels). Values of central source luminosity are marked within the panels.  
We present results for the gas located at $24$~pc from M60-UCD1 (upper row of panels) and 
at $0.2$~pc (bottom row of panels). The luminosity of the NSC is equal to
log($L_{\rm stars}$/erg~s$^{-1})=40.66$ at 24 pc and log($L_{\rm stars}$/erg~s$^{-1})=38.47$ at 0.2 pc, 
and the volume mechanical heating is always $H_{\rm ext}=3.1 \times 10^{-24}$~erg~s$^{-1}$~cm$^{-3}$.}
\label{fig:UCD1_equilibrium}
\end{figure*}

 Stars in M60-UCD1 are much older than stars in the Sgr A*
neighborhood, so they are not sources of vigorous winds. Main sequence
stars of the age of the Sun are sources of fast winds.  The solar wind
has a wind outflow rate of $\sim 10^{12}$ g s$^{-1}$,  and an outflow velocity
$\sim 500$ km s$^{-1}$.  If we assume
that all the stars are in this form and that they uniformly occupy the core within a radius of 
24 pc, we arrive at a main sequence stellar mass loss of $3 \times 10^{-6} M_{\odot}$ yr$^{-1}$ and their contribution to the total heating
\begin{equation}
H_{\rm MS} = 1.0 \times 10^{-25} {\rm erg \;s}^{-1} {\rm cm}^{-3}\;{\rm for}\; r < 24\; {\rm pc}.
\end{equation}
 Most of the mass loss, however, happens at the post main sequence stage. There, the mass loss is driven by
 the presence of dust in the stellar envelope. The total mass loss of a single star is of order of $10^{-6}$ M$_{\odot}$ yr$^{-1}$, but the strong wind stage is short, and the wind velocity is small, of
order of 10 km s$^{-1}$. This velocity is much smaller than the stellar velocity dispersion in  UCD1
(70 km s$^{-1}$, Strader et al. 2013). The effective mass loss from the entire stellar population has been
 estimated to be
$2.5 \times 10^{-12}$ M$_{\odot}$ yr$^{-1}$ M$_{\odot}^{-1}$, under  the assumption of a Salpeter mass
distribution \citep{pellegrini2011}.  Taking this mass loss, and the dispersion velocity instead of the
wind velocity we obtain the total heating
\begin{equation}
H_{\rm PMS} = 7.0 \times 10^{-25} {\rm erg \;s}^{-1} {\rm cm}^{-3}\;{\rm for}\; r < 24\; {\rm pc}.
\end{equation}
Although supernova eruptions in such a small and old stellar population are rare, their average 
contribution is important. Assuming the intrinsic B band luminosity $6 \times 10^5 L_{\odot}$ and using Eq.~3
of \citet{pellegrini2011}, we obtain the
mean energy deposition per unit volume
\begin{equation}
H_{\rm SN} = 2.7 \times 10^{-24} {\rm erg \;s}^{-1} {\rm cm}^{-3}\;{\rm for}\; r < 24\; {\rm pc}.
\end{equation}

We take the sum of the contributions from the three stellar heating chanels as the final 
stellar heating rate:
\begin{equation}
H_{\rm ext} = 3.1 \times 10^{-24} {\rm erg \;s}^{-1} {\rm cm}^{-3}\;{\rm for}\; r < 24\; {\rm pc}.
\label{eq:UCD1_heating}
\end{equation}
This is more than 6 orders of magnitude lower than for Sgr A* but comparable to various heating processes
 operating in the disk of the Milky Way galaxy \citep{klessen2014}.


We calculate the thermal equilibrium curves at two distances from the centre, i.e. at 24~pc (which is 
the half-light radius of the NSC), and further inside at 0.2~pc. The gas is expected to be less
 dense, therefore we consider clouds spanning the range from $\log n_{\rm H} = -3$  
 up to $\log n_{\rm H} = 2.5$. The results for both distances are  shown in 
 Fig.~\ref{fig:UCD1_equilibrium} upper and lower panels, respectively. 
The upper panels clearly show that the distance of 24~pc from the center is far enough to prevent 
matter from being strongly ionized. The radiation is not hard enough to heat up the 
gas to the high equilibrium branch. 
High (unrealistic) temperatures appear for very low densities (below 0.03~cm$^{-3}$) only when 
mechanical heating is included (right panel). 
However at a distance of 0.2~pc from the center, 
the radiation field from the accreting gas is hard enough for the high temperature branch to appear. Soft photons from old stars cool down this gas, which is clearly shown in the middle bottom panel. 
Finally, closer to the center 
mechanical heating is not able to heat the gas to such unrealistically high temperatures.

\begin{figure*}
\includegraphics[scale=0.45]{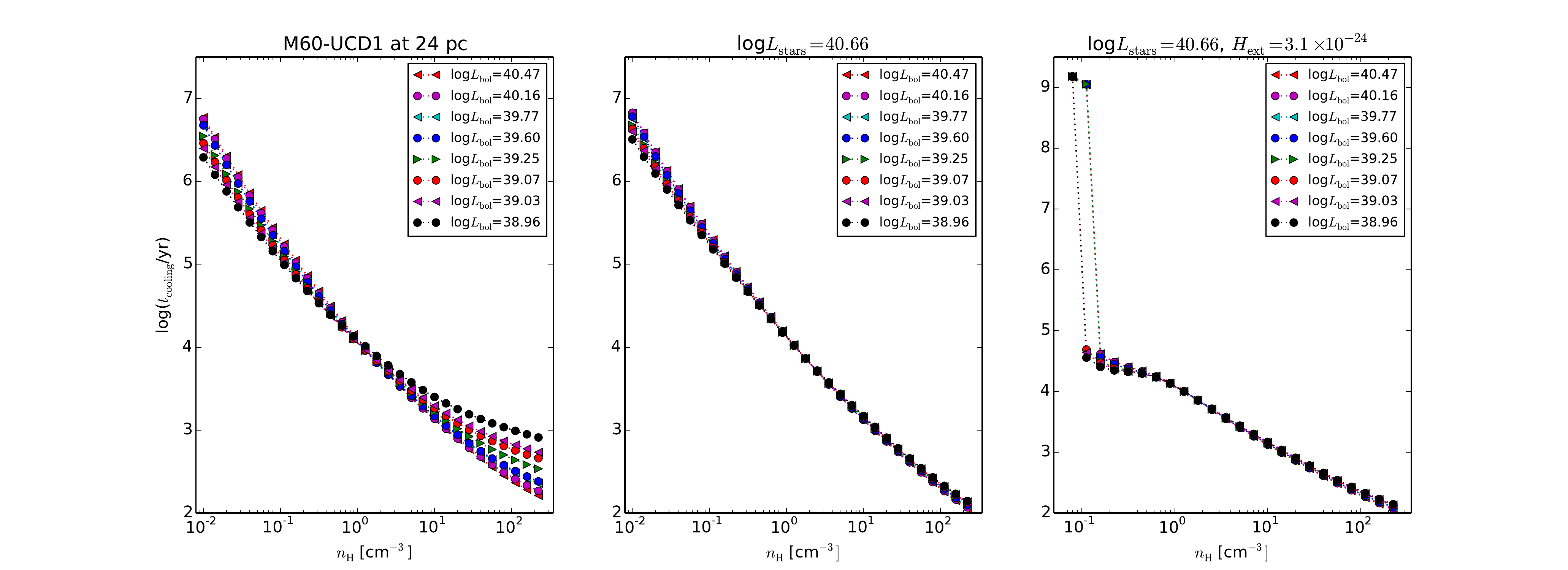}
\includegraphics[scale=0.45]{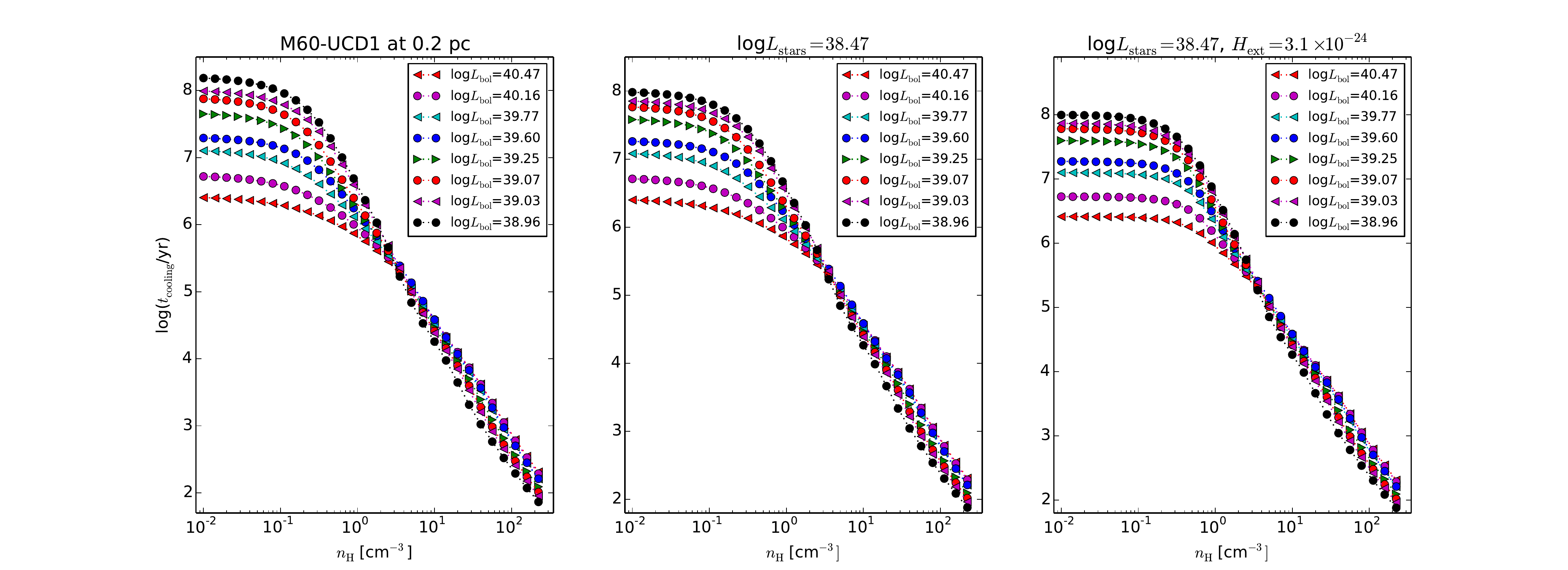} 
\caption{Equilibrium cooling timescale in case of  M60-UCD1, for different luminosity states of the 
radiation: from the central source  only (left panels), 
together with heating by stellar radiation (middle panels), and together with mechanical heating 
by winds (right panels). 
The results presented here are for the same model parameters as in Fig.~\ref{fig:UCD1_cooling}.}
\label{fig:UCD1_cooling}
\end{figure*}

The thermal timescales for the plasma in equilibrium are shown in Fig.~\ref{fig:UCD1_cooling}. 
The timescales are shorter for the
cold dense gas, but they are still quite long, of the order of $10^7$~yr, for the low density high temperature plasma located at 0.2~pc. In the case when mechanical heating dominates in the gas at the 
distance of $24$~pc, the cooling timescale for low density plasma can reach $10^{10}$~yr.
 
\begin{figure}
\includegraphics[scale=0.51]{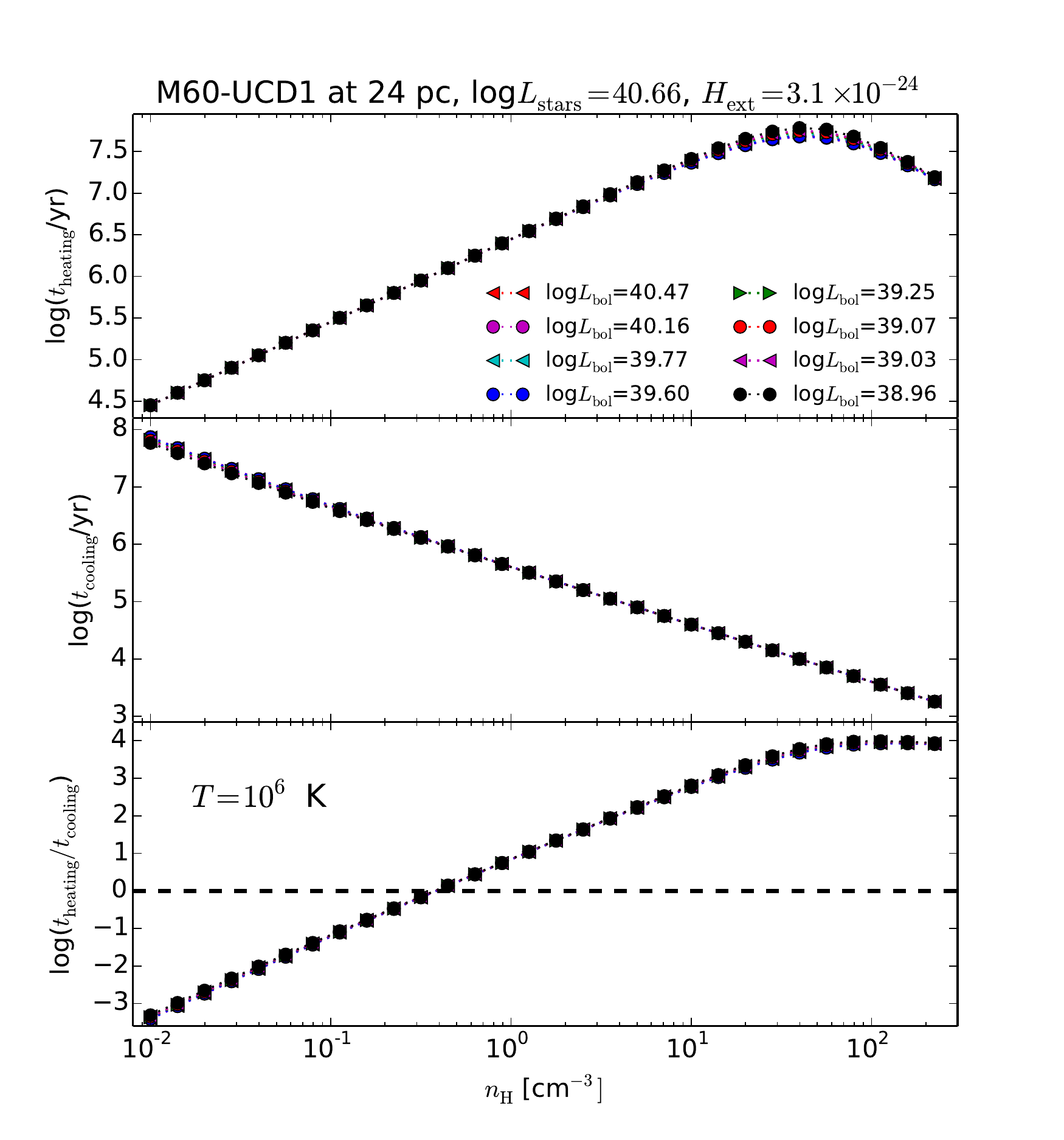} 
\caption{M60-UCD1 heating (upper panel) and cooling (middle panel) time-scales plotted against cloud 
density. Bottom panel represents ratio of those two values.  The ISM gas in each case is
heated by radiation from the center with different luminosity states,  radiation from stars 
of log($L_{\rm stars}$/erg~s$^{-1}) = 40.66$, and volume mechanical heating
 $H_{\rm ext} = 8.8 \times 10^{-26}$  erg s$^{-1}$ cm$^{-3}$.
Each cloud is assumed to have a temperature $T=10^6$ K and is located at a distance of 24~pc from 
the center.}
\label{fig:ucdthermal24}
\end{figure}

We can estimate the mean density of the hot phase originating from the thermalized stellar winds 
as the basis of the observed X-ray emission ($L_{\rm X} = 1.3 \times 10^{38}$ erg s$^{-1}$),
 assuming this  emission  is predominantly due to bremsstrahlung.
 We adopt a stellar wind temperature, $T_{\rm wind} = 10^6 $ K. Assuming the size of the emitting 
region to be the nuclear radii, we have
\begin{equation}
n_{\rm mean} = 7.32  \times \left({T \over 10^6 {\rm K}}\right)^{-1/4} {\rm cm}^{-3}.
\end{equation}
For these densities, the timescales in equilibrium are quite short, $10^3$~yr, and the 
plasma moves toward the equilibrium curve shown in Fig.~\ref{fig:UCD1_equilibrium}. 
However, at this density and temperature the plasma is not in local thermal equilibrium, 
as can be seen from the equilibrium curve. The cooling dominates,
 since the mean bremsstrahlung emission is of order of $7.64 \times 10^{-23}$ erg s$^{-1}$~cm$^{-3}$
(i.e.\ almost two orders of magnitude higher than the heat input from stellar winds). 
Therefore, we perform computations of the cooling timescale
for a fixed temperature and a range of densities to see the trend in the plasma far from thermal 
equilibrium. Plasma produced from stellar winds cools down if the density is higher than 
0.1~cm$^{-3}$ (see Fig.~\ref{fig:ucdthermal24}), and at the mean density estimated above it 
needs $\sim 10^4$~yr to cool down.

If the star  is close to the edge of the nucleus and the outflow velocity is of the same order as
 the wind velocity, the plasma may leave the nucleus before it cools significantly.  The crossing 
 time is only
\begin{equation}
t_{\rm cross} = 4.7 \times 10^4 [{\rm yr}].
\end{equation} 
 The plasma in outer parts of the nucleus thus preserves the energy content formed in the 
stellar wind interactions. An adiabatic description offers an acceptable approximation although 
cooling is not completely negligible.

However, if the star is located close to the stagnation point between the outflow and inflow, 
then the velocity is much lower, the plasma can cool down and form colder and denser clumps
\citep[see][for a discussion in the context of Sgr~A*]{quataert2004,shcherbakov2010}. 

The stagnation radius, $R_{\rm st}$, for M60-UCD1 can be estimated using the formula \citep{silich2008}
\begin{equation}
R_{\rm st} = \left({GM_{\bullet} \over 2 R_{\rm NSC} v_{A\infty}^ 2}\right)^{2/3} R_{\rm NSC},
\end{equation}
where $R_{\rm NSC}$ is the NSC (half-light) radius, and $v_{A\infty}$ is the terminal velocity of 
the stellar wind. Here we neglected dimensionless coefficients which depend on the polytropic index.
 Assuming a terminal velocity three times  higher than the wind velocity (i.e. 1500 km~s$^{-1}$) 
 we obtain  $R_{\rm st} = 0.01 R_{\rm NSC}$, deep inside the NSC.  Most of the
material thus starts to flow out.  The radiative cooling is then partially assisted by the adiabatic 
cooling. The cooling timescale is an order of magnitude longer than in the outskirts of the galaxy 
(see Fig.~\ref{fig:ucdthermal02})  but the cooling still dominates and adiabatic expansion will 
assist this trend. Therefore, spontaneous cloud formation is likely, although time-dependent 
modeling coupled with the inflow/outflow dynamics would be necessary to confirm this trend.

This picture is strongly different from the situation in Sgr~A*, where the locally inserted stellar
wind cannot cool down due to the stellar heating always dominating the radiative cooling inside 
the inner parsec.

\begin{figure}
\includegraphics[scale=0.51]{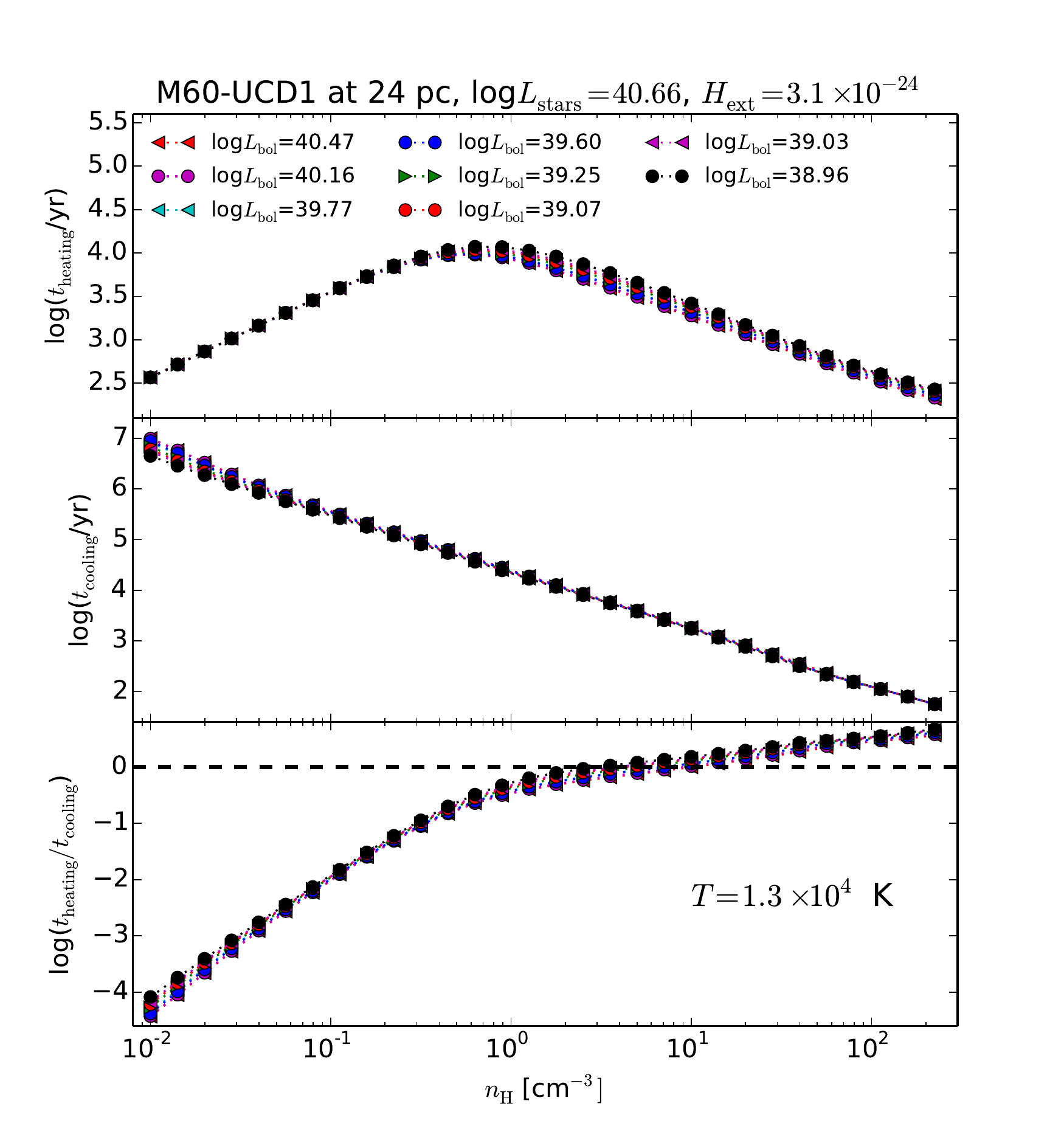} 
\caption{The same as in Fig.~\ref{fig:ucdthermal24} but each cloud is assumed to have 
temperature $T=1.3 \times 10^4$~K.}
\label{fig:ucd4thermal24}
\end{figure}

\begin{figure}
\includegraphics[scale=0.51]{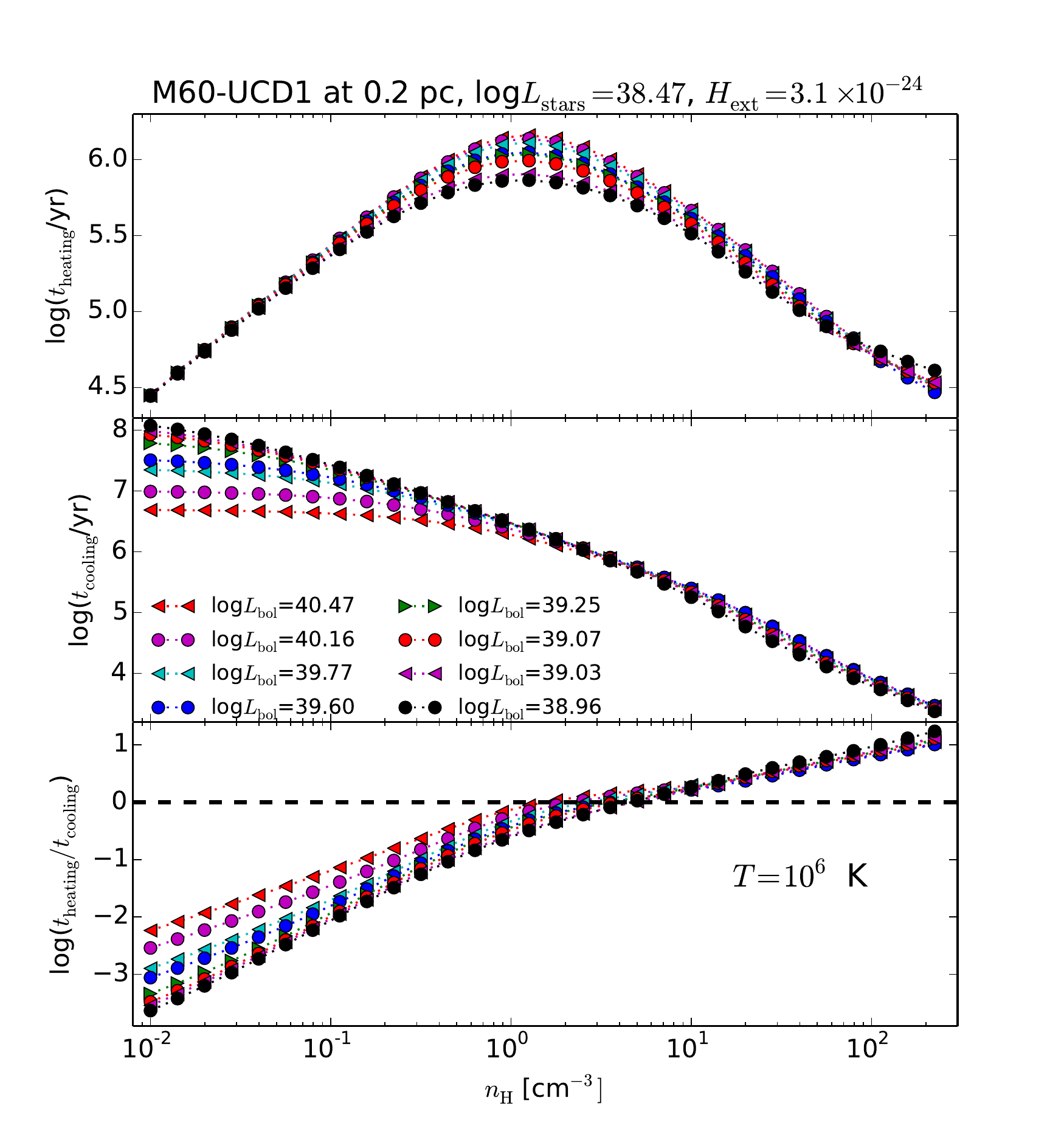} 
\caption{The same as in Fig.~\ref{fig:ucdthermal24} but for gas located at 0.2~pc from the
M60-UCD1 center.}
\label{fig:ucdthermal02}
\end{figure}

\begin{figure}
\includegraphics[scale=0.51]{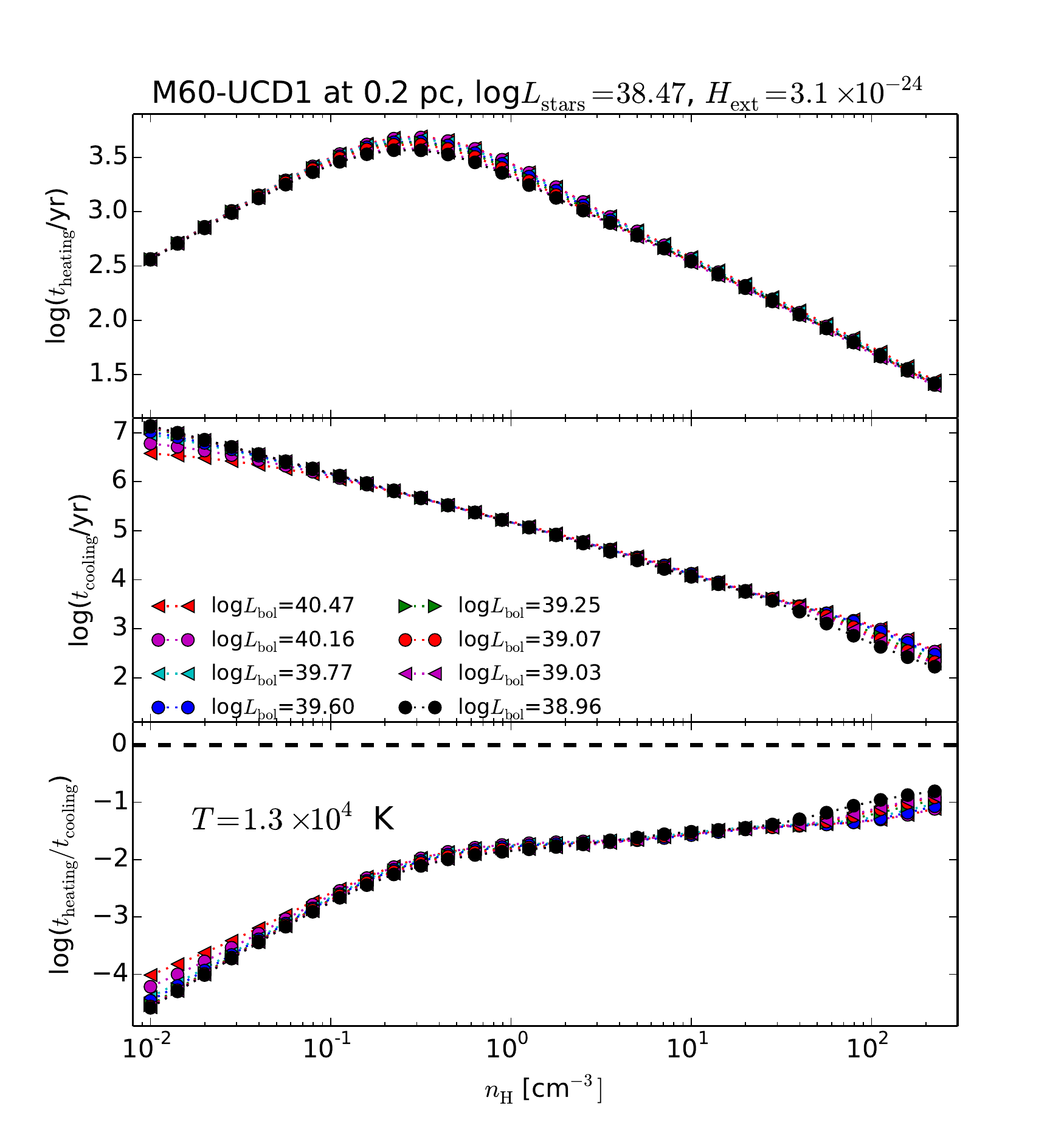} 
\caption{The same as in Fig.~\ref{fig:ucdthermal02} but each cloud is assumed to have 
temperature $T=1.3 \times 10^4$~K.}
\label{fig:ucd4thermal02}
\end{figure}

\section{Discussion}
\label{sec:dis}

We considered the thermal stability of plasma in the innermost regions of
compact stellar clusters surrounding a supermassive black hole.
 We discussed in particular two cases: the nuclear cluster 
in the Milky Way and the Ultra-Compact Dwarf galaxy M60-UCD1. We found that the conditions are more 
extreme in the case of Sgr~A* due to the higher compactness of the nuclear cluster: stellar 
heating prevents the spontaneous formation of cold clouds out of the freshly inserted stellar 
wind plasma. On the other hand, if cold dense clumps of material happen to be in this region,
 they can stay in thermal equilibrium. From observations, the existence of cold plasma in the mini-spiral
  region is well known which is consistent with our scenario but the 
 origin of that plasma is then 
  clearly non-local.  Fresh plasma from stellar winds joins the hot X-ray emitting phase seen in 
  X-ray images and it forms an outflow, as modeled by \citet{quataert2004} and \citet{shcherbakov2010}.  

In the case of M60-UCD1 mechanical heating by stellar winds is not as intense and we expect
 spontaneous formation of cold clouds in the inner part of the galaxy, close to the stagnation 
 radius, since the cooling timescales are then shorter than the inflow/outflow timescale. 
Let us note that our paper addresses only the direction of time evolution in case  of a
 thermal imbalance in the medium, leaving aside the dynamical interaction of the cold clouds
  with the hot surrounding medium.

A complete non-stationary picture of the behavior of the multi-phase medium can be 
obtained only by three-dimensional time-dependent numerical 
computations. The main advantage of the numerical approach is obviously the consistency 
of the description and the possibility to include additional 
effects. On the other hand, a large number of free parameters complicates the interpretation, 
and so both the numerical and semi-analytical approaches 
are useful and complement each other. 

It is interesting to note that, recently, \citet{barai12} have shown  that 
cooler clumps form filaments near a SMBH, which then 
accrete  faster than the surrounding hot plasma. \citet{hueyotl2013} studied the evolution 
of matter reinserted by stars in a rotating 
young NSC. These authors employed a hydrodynamic model that includes gravitational attraction 
from the central SMBH and NSC, and it also takes radiative cooling and heating from the central
source into account. By following the formation of thermally unstable zones 
for a large set of parameters of the NSC and SMBH, including those relevant for the present 
state of the Galactic centre, \citet{hueyotl2013} 
concluded that for a NSC of mass $3.3\times10^7\,M_{\odot}$ and size 10\,pc, SMBH mass 
$M_\bullet\simeq1\times10^6\,M_{\odot}$ and 
accretion luminosity $L\simeq1.3\times10^{38}$\,ergs\,s$^{-1}$, the central source of radiation 
heats the gas up to a few $\times10^7$\,K. This prevents the development of 
the Thermal Instability. 

In the context of galaxy outflows, \citet{scannapieco2015} have recently performed numerical 
simulations of the evolution of cold clumps embedded 
in an outflowing medium. Their results show that clouds can survive within a supersonic flow 
for a period of time long enough to be accelerated 
by the flow, although the outflowing clouds are gradually destroyed. In our model 
we consider a close region above and below the stagnation 
radius, and so the orientation of the radial motion can be directed with either positive or 
negative radial velocity component, depending on the interplay of hydrodynamic 
friction by the ambient medium, radiation pressure from stars in the NSC, and the
 gravitational attraction of the SMBH.

In our considerations we assumed that the electrons and ions are well coupled and form a 
single-temperature plasma. This is justified since several studies of spherical hot accretion 
onto Galactic Center show that 
a two-temperature plasma develops only within the innermost tens of gravitational radii (<0.1 arcsec) from 
the black hole \citep{moscibrodzka2006,moscibrodzka2009,roman2014}. In addition, strong arguments
in favour of very efficient electron-ion coupling \citep{bisnovatyi1997} imply that this region is very small.

We have not yet included the role of magnetic fields in our scheme but we note that  they are important for TI, although this issue is difficult to explore due to the enormous richness of possible configurations.  Already in his seminal paper, \citet{field1965} realized that magnetic fields penetrate the interstellar plasma  and influence the onset of TI. First, the internal pressure 
 gets increased by the magnetic term which must be added to the gas pressure \citep{langer1978}.
  This is, however, a rather minor and mainly quantitative change under usual conditions of 
the interstellar medium, unless the magnetic field  exceeds the equipartition value
($P_{\rm m}\gtrsim P_{\rm g}$). This may be the case very close to the horizon, where Sgr~A* 
flares are thought to occur.  A more important mechanism is anisotropic transfer
 of heat across the field lines, for which the presence of a magnetic 
field of relatively weak (sub-equipartition, $P_{\rm m}\ll P_{\rm g}$) intensity is sufficient. 
This is caused by the fact that the motion of electrons is greatly reduced in 
the direction across the field lines. However, in the plane-parallel approximation the growth 
timescale of the  condensation instability is not diminished very strongly, 
unless the magnetic field is strictly perpendicular to the temperature gradient \citep{hoven1984}. 

\citet{celotti1999} studied the possibility that cold material may be confined by
 the ambient magnetic 
field that pervades the hot medium, thereby establishing conditions for co-existence of phases
 at very different temperature. These authors considered a quasi-equipartition
case when the magnetic pressure approximately equals the gas term. According to their scenario the 
main role of the magnetic field is  to trap plasma in clouds that have enough time to cool down to 
the Compton temperature. Quite recently, \citet{Wang2012} explored 
the velocity dispersion of cold clouds embedded in an advection dominated accretion flow (ADAF), and 
\citet{Khajenabi2015}  also included the effect of weak (sub-equipartition) magnetic fields. 
While the cold phase can naturally attain the form of localized clouds, 
 one-dimensional (filamentary) structures have also been suggested.

The role of magnetically induced anisotropy depends critically on the magnetic field 
structure (the tangled small-scale magnetic loops vs.\ large-scale organized field lines) and the 
ionization state of the medium (hot, fully ionized plasma is more sensitive to
the imposed magnetization than the cooler medium with only partial degree of ionization).
 \citet{burkert2000} found that tangled magnetic fields reduce the conductive heat flux. 
 As a result, low-amplitude fluctuations can grow until the nonlinear effects start
operating at length-scales of order $\sim0.01$~pc. Magnetization enhances the filamentary 
structures, which are inherently three-dimensional and their discussion requires numerical approaches 
\citep{parrish2009,mccourt2012,sharma2012}. 

Thus, while the additional mechanical heating from stellar winds tends to suppress the Thermal 
Instability,  magnetic fields effectively insulate the gas elements in the direction 
perpendicular to the field lines, thereby decreasing the heat conduction and permitting large 
temperature gradients across the field lines. As a consequence, the multi-phase structure is maintained. Nonetheless, the magnetic structure 
and the source of heating are critical but largely unknown.

\citet{tenorio-tagle2013} studied gas and dust cooling in centers of young, massive, compact star 
clusters (typical length-scales of $\sim10$~pc). Interestingly, TI  operates and 
an unstable region develops also in these systems. Based on {\sc Starburst99} \citep{leitherer1999} 
synthesis computations we find that matter reinserted in the thermally unstable volume is 
reprocessed into dense clouds and there it leads to a new phase of star 
formation.\footnote{Let us note that, similarly to our analysis in the previous section but without 
the central black hole, \citet{tenorio-tagle2013} found that only a limited fraction of the released 
mechanical energy from stars is transferred back into the ISM.} Very recently,
\citet{dale15} reviewed the operation of TI  in the context of state-of-art
 numerical simulations of the multi-phase environment in stellar systems with feedback.
 In fact, the physical origin of feedback processes
in stellar systems is analogous to the widely discussed feedback mechanism in active galactic nuclei
 \citep{fabian12}, although the scales and conditions are vastly different. An important message
  from these studies is the recognition of the fact that these systems behave 
in a self-regulating regime for which an accurate description of cooling and heating processes is 
essential.

Additional mechanisms contribute to the feedback processes and may eventually enhance the mechanical heating above the level assumed in our simplified scenario, but only some of them can be relevant in evolved stellar systems \citep{ciotti2007,pellegrini2012}. 
In particular, after a star-forming burst at a certain moment, the mass and energy injection from 
massive stars peak at earlier time with respect to the input from the post-main-sequence population, and from supernovae Type II explosions.

\section{Conclusions}
\label{sec:con}
We studied the conditions for the onset of the Thermal Instability in compact stellar systems with 
a supermassive black hole residing in the  core. As the Thermal Instability develops, it can 
help cold clumps to survive in the surrounding hot medium and drive them toward the central SMBH, thus enhancing the mass accretion rate during episodes of clumps disruption and inflow.

The Nuclear Star Cluster in the centre of the Milky Way and the Ultra-Compact Dwarf galaxy UCD1 
near M60 galaxy represent prototypical systems with a small half-light radius and large mass-to-light 
ratio, suggesting that an interplay between  gravitational, radiative,
and hydrodynamic influences leads to an interesting richness of the evolutionary tracks of the
 accreting SMBH. A large number of UCDs have been discovered in the last decade, 
however the presence of SMBHs at the centres of UCDs remains an open question. 
Sgr~A* is the best-resolved extreme example of a  galactic nucleus with a Nuclear Star Cluster 
in addition to a SMBH.
 
 To explore the possibility of TI as a driver
that triggers variability via accretion episodes, we discussed an appropriate 
definition of the ionization parameter $\Xi$  and studied the position of different
systems with respect to the S-curve in the instability diagram. In addition to the effect of X-ray 
irradiation from the core we also took the cooling effect of ambient stellar 
radiation  and mechanical heating from colliding winds into 
account, and we considered the presence of stagnation radius $R_{\rm st}$ in the 
Bondi-like hot inflow/outflow. We showed that Thermal Instability indeed operates 
under suitable conditions while it can be suppressed in other parts of the parameter space
during the evolution of the system. Cold clouds can thus remain within the surrounding hot diluted 
medium for a relatively long period.

A more complete multi-wavelength picture of these systems can help to constrain the state of
 their ISM in the future. Current evidence for the presence of a SMBH at the centre of M60-UCD1 
 arises from dynamical modeling of adaptive optics kinematic data \citep{seth2014}. 
 However, apart from a tentative detection in X-rays, there is very little information 
 about the SED of the central black hole. 
As we have seen in Section 3.2, the shape of the SED of the central black hole influences the 
formation of TI close to the SMBH.
Follow-up observations in the X-ray and radio continuum can help to detect signatures of 
accretion in the central black hole and provide a more realistic SED. 
Observations in the infrared can also reveal the presence of dust/gas in the central 
regions to get a better estimate of the properties of the ISM in those objects.

\section*{Acknowledgments}
This research was supported by  Polish National Science Center grants No. 2011/03/B/ST9/03281, 
 2013/10/M/ST9/00729,  2015/17/B/ST9/03422, 2015/18/M/ST9/00541 and by 
Ministry of Science and Higher Education grant W30/7.PR/2013. We also received funding from the 
European Union Seventh Framework 
Program (FP7/2007-2013) under grant agreement No. 312789, and Czech Science Foundation 
``Albert Einstein Center for
Gravitation and Astrophysics''  (GA\v{C}R 14-37086G). We acknowledge the Czech Ministry of Education,
 Youth and Sports  COST project 
LD15061 titled ``Astrophysics of toroidal fluid structures around compact stars''.

\bibliographystyle{mnras}
\bibliography{refs}

\end{document}